\documentclass[letterpaper,twocolumn,10pt]{article}
\usepackage{usenix}

\usepackage{tikz}
\usepackage{amsmath}

\usepackage{graphicx}
\usepackage{epsfig,endnotes}
\usepackage{subcaption, float, soul}
\usepackage{array}
\usepackage{multirow}
\usepackage{xspace}
\usepackage{verbatim}
\usepackage{wrapfig}
\usepackage[english]{babel}
\usepackage{bbding}
\usepackage{pifont}
\usepackage{wasysym}
\usepackage{caption}
\usepackage[shortcuts]{extdash}  % make latex hypheneate even if a word contains a dash. Use '\-/' instead of '-'.
\usepackage{siunitx}
\usepackage[nopostdot,style=super,toc,acronyms,nonumberlist]{glossaries}
\usepackage[ruled,vlined]{algorithm2e}
\usepackage{algpseudocode}
\usepackage[normalem]{ulem} % in this directory
\usepackage{tabularx, url, booktabs} %
\usepackage{xurl}
\usepackage{listings}
\usepackage{circledsteps}
\usepackage{csquotes}
\usepackage{color,colortbl}
\usepackage{pifont}
\usepackage{varwidth}
\usepackage{grumble}
\usepackage{enumitem}
\usepackage{hyperref}
\usepackage{bm}
\usepackage{stfloats}
\usepackage{makecell}

\usepackage[mono=false]{libertine}
\usepackage{libertinust1math}
\usepackage{ascii}

\newboolean{revision}
\setboolean{revision}{false}
\definecolor{BLUE}{RGB}{0,0,255}
\ifthenelse{\boolean{revision}}{
    \newcommand{\revision}[1]{\textcolor{blue}{#1}}
}{
    \newcommand{\revision}[1]{#1}
}

\ifthenelse{\boolean{revision}}{
    \newcommand{\revisionmyparagraph}[1]{\myparagraph{\textcolor{blue}{#1}}}
}{
    \newcommand{\revisionmyparagraph}[1]{{\myparagraph{{#1}}}}
}

\newcommand{\sysnamenormal}{Wallet}
\newcommand{\sysname}{\textsc{\sysnamenormal}\xspace}
\newcommand{\projectname}{\textsc{\sysnamenormal}\xspace}

\newcommand{\largecomment}[1]{}

\newcommand{\boundedbox}[1]{
    \noindent\fbox{%
        \parbox{0.46\textwidth}{%
        #1
        }
    }
}

\newcommand{\tabfontsize}{\fontsize{8}{9}\selectfont}

\newcommand{\myparagraph}[1]{\noindent\textit{\textbf{{#1}.}~}}
\newcommand{\out}[1] {}

\newcounter{codeLineCntr}

\reversemarginpar
\newif\ifnotes
\notestrue

\newcommand{\punt}[1]{}

\renewcommand{\eqref}[1]{Equation~(\ref{eq:#1})}

\newcommand{\proc}[1]{\ifmmode\mbox{\textsc{#1}}\else\textsc{#1}\fi}

  \newcommand{\func}[1]{\ifmmode\mathrm{#1}\else\textrm{#1}fi} %
\newcounter{remark}[section]

\begin{document}
%-------------------------------------------------------------------------------

%don't want date printed
\date{}

\title{Confidential Serverless Computing}
\author{
{\rm Patrick Sabanic} \quad {\rm Masanori Misono} \quad {\rm Teofil Bodea} \quad {\rm Julian Pritzi} \quad \\ {\rm Michael Hackl} \quad {\rm Dimitrios Stavrakakis} \quad {\rm Pramod Bhatotia} \\
Technical University of Munich
}

%-------------------------------------------------------------------------------

%-------------------------------------------------------------------------------
%% COPYRIGHT REMOVAL
% \setcopyright{none}
% \settopmatter{printacmref=false} % Removes citation information below abstract
% %% FOOTNOTE REMOVAL
% \renewcommand\footnotetextcopyrightpermission[1]{} % removes footnote with conference information in first column
% %% PAGE NUMBERING
% \settopmatter{printfolios=true} % add page numbering

\maketitle
\pagestyle{plain} % page numbering

% \balance

%-------------------------------------------------------------------------------
\begin{abstract}
Although serverless computing offers compelling cost and deployment simplicity advantages, a significant challenge remains in securely managing sensitive data as it flows through {\em the network of ephemeral function executions} in serverless computing environments within untrusted clouds. While Confidential Virtual Machines (CVMs) offer a promising secure execution environment, their integration with serverless architectures currently faces fundamental limitations in key areas: \emph{security}, \emph{performance}, and \emph{resource efficiency}. 

%Serverless computing offers cost-effective and simplified deployment, but a significant hurdle remains in securely managing sensitive data within untrusted cloud infrastructures. While Confidential Virtual Machines (CVMs) provide a promising secure execution substrate to address the security challenge, their integration with serverless architectures presents limitations in \emph{security}, \emph{performance}, and \emph{resource efficiency}. 

We present \projectname{}, a lightweight confidential computing system for secure serverless deployments. By employing nested confidential execution and a decoupled guest OS within CVMs, \projectname{} runs each function in a minimal ``trustlet'', significantly improving security through a reduced Trusted Computing Base (TCB). Furthermore, by leveraging a data-centric I/O architecture built upon a lightweight LibOS, \projectname{} optimizes network communication to address performance and resource efficiency challenges.

%, focusing on security, performance, and resource efficiency,
Our evaluation shows that compared to CVM-based deployments, \projectname{} has a 4.3$\times$ smaller TCB, improves end-to-end latency (15–93\%), achieves higher function density (up to 907$\times$), and reduces inter-function communication (up to 27$\times$) and function chaining latency (16.7-30.2$\times$); thus, \projectname{} offers a practical system design for confidential serverless computing.

\end{abstract}

\section{Introduction}
\label{sec:introduction}

% \myparagraph{Context and motivation} 
% \nuno{I don't think the paragraph headings are necessary; the intro is very well structured, and dropping them can save a few precious lines.}
%Serverless computing is a rapidly evolving cloud paradigm offering low costs, automatic scaling, and simplified deployment~\cite{serverless-growth}. Major cloud providers now offer serverless solutions~\cite{aws-lambda, azure-functions, ibm-serverless, google-cloud-functions, cloudflare-workers}. 

Serverless computing is a rapidly evolving cloud paradigm offering low costs and simplified deployment~\cite{serverless-growth}. %, with major cloud providers now offering serverless solutions~\cite{aws-lambda, azure-functions, google-cloud-functions}.
As serverless applications increasingly manage sensitive data, particularly in AI/ML, strong security measures are of utmost importance~\cite{serverless_on_ML, serverless_inference}. 
%Securely managing data becomes particularly challenging in serverless environments where short-lived, distributed functions are chained through network communication~\cite{citeme}.
However, securely managing data within serverless computing is particularly challenging due to their ephemeral and distributed execution, where short-lived functions are chained through network communication~\cite{secure_serverless, clemmys, kalium}.

To this end, Confidential Virtual Machines (CVMs)~\cite{tdx, sev, cca} provide a promising approach for the secure execution of serverless workloads in untrusted cloud environments~\cite{confidential_computing}, as they protect the confidentiality and integrity of entire VMs and do not require significant application modifications~\cite{sgx, trustzone}. %, unlike process-based solutions~\cite{sgx, trustzone}. 
However, due to strict serverless computing requirements, simply deploying serverless workloads directly within CVMs reveals critical inherent limitations (\autoref{sec:motivation}). In particular, despite its portability benefits, this approach exhibits \textit{non-ideal security properties}, incurs \textit{prohibitive performance overheads}, and leads to \textit{costly resource inefficiencies}.
In terms of security, CVMs rely on a full guest OS to host serverless applications, resulting in a bloated Trusted Computing Base (TCB) that expands the attack surface due to potential vulnerabilities in the untrusted OS (\autoref{subsec:mot:tcb}). From a performance perspective, CVMs suffer from long boot times---often exceeding function execution duration---due to strict security mechanisms (e.g., memory encryption) (\autoref{subsec:mot:boot_times}). They also introduce significant I/O communication overheads, which are particularly detrimental in function chaining scenarios (\autoref{subsec:mot:network_communication}) and require costly per-instance attestation, adding considerable latency to short-lived executions (\autoref{subsec:mot:attestation}). Finally, resource efficiency is severely impacted as CVM-based serverless frameworks struggle with function scheduling (\autoref{subsec:mot:scheduling}), primarily due to consolidation challenges stemming from ineffective memory deduplication and hardware density constraints (\autoref{subsec:mot:server_consolidatiion}).

Despite the advancements in virtualization architectures, \textit{no existing approach can fully resolve the combined challenges of security, performance, and resource efficiency in confidential serverless computing}. 
Traditional VMs provide hardware-based isolation but incur substantial boot delays~\cite{vm_startup_perf}, high inter-VM communication costs~\cite{xenloop}, and maintain large TCBs because of their full virtualization stacks. 
Lightweight VMs (e.g., FireCracker~\cite{firecracker}, Kata~\cite{kata}) optimize function density and startup latency~\cite{LightVM}, and reduce TCBs by trimming hypervisor and guest OS components~\cite{firecracker}; however, they still suffer from significant I/O overheads and designs that complicate security verification~\cite{hyperlight, serverlessCoCo, confidential-containers}. 
Specialized OS architectures (e.g., LibOSes~\cite{cubicleOS, gramine1, kuvaiskii2024gramine}, Unikernels~\cite{unikraft, unikernels, uio2024}) deliver near-instant startups by compiling applications with minimal kernel libraries but require developer efforts to craft application-specific images and sacrifice compatibility. 
CVMs (e.g., AMD SEV-SNP~\cite{sev-snp}, Intel TDX~\cite{tdx}, ARM CCA~\cite{cca}), in turn, provide hardware-based security properties but introduce significant network communication overheads~\cite{li_bifrost_2023,bounce_buffer,misono2024confidential}. % (e.g., bounce buffers~\cite{bounce_buffer}). 
In summary, while each approach addresses individual aspects of the problem, a holistic solution that reconciles all these concerns remains an open challenge.

As a solution, we propose \projectname{}, a lightweight confidential computing system that enables secure serverless deployments in untrusted clouds by fundamentally rethinking how sensitive workloads are isolated and executed inside CVMs. In \projectname{}, each function runs in its own \emph{trustlet}---a minimal serverless process that executes within a secure environment inside a CVM. With this design, client requests are securely routed through the untrusted guest OS, while the actual execution occurs in trustlets that encapsulate only the critical application code and the minimal system support required to serve these functions. This arrangement leverages two key mechanisms: \emph{nested confidential execution} and a \emph{decoupled guest OS architecture}. Inspired by nested virtualization~\cite{ben-yehuda_turtles_2010} and ARM’s TrustZone model~\cite{trustzone, tlr}, we  partition the CVM space into smaller nested isolated regions. By confining sensitive code to trustlets, \projectname{} significantly reduces the overall TCB and minimizes the attack surface, thereby overcoming the limitations of hosting workloads on a monolithic guest OS.

To achieve high performance and fast function instantiation, \projectname{} adopts several techniques. First, it employs pre-initialized \emph{process templates} (inspired by Android’s zygote model~\cite{zygote, lee_morula_2014}) to expedite the bootstrapping of trustlets. Complementing this, a \emph{dynamically loadable LibOS} enables trustlets to achieve the minimal startup latency characteristic of unikernel-based systems~\cite{unikernels, ukl, kylinx} while retaining flexibility for diverse workloads. 
Moreover, \projectname{} introduces a \emph{data-centric I/O architecture} that leverages high-density function co-location and replaces conventional CVM networking for inter-function communication, optimizing communication paths via secure \emph{data objects}. It also incorporates \emph{differential attestation}-an innovative technique that reuses pre-measured components to reduce attestation latency by focusing on only the mutable parts during invocation.
% Moreover, \projectname{} introduces a \emph{hybrid I/O model} that optimizes function communication paths via secure shared-memory channels and incorporates \emph{differential attestation}—an innovative technique that reuses pre-measured components to reduce attestation latency by focusing on only the mutable parts during invocation.

Collectively, these mechanisms enhance resource efficiency. As process templates and the LibOS are designed to share memory, and trustlets are ephemeral---occupying resources only during function execution---\projectname{} achieves a higher density of function deployments per CVM. This tight integration of performance optimizations with a thin TCB not only accelerates startup and execution but also allows for more efficient resource utilization, thereby supporting a larger number of concurrent function requests within the same memory footprint.

We implement \projectname{} on AMD SEV-SNP~\cite{sev-snp} with source code and evaluation setup to be \textbf{publicly available}.
% Our evaluation shows that, compared to CVMs, \projectname{} reduces the TCB by 4.3$\times$, while in real serverless applications, it achieves a lower end-to-end latency in both cold (86.01-94.46\%) and warm (15.14\%) start cases.
Our evaluation shows \projectname{} reduces TCB by 4.3$\times$ compared to CVMs while achieving lower end-to-end latency in real serverless applications for both cold (85.10-93.44\%) and warm (14.97\%) starts.
% Further, \projectname{} consistently demonstrates performance improvements in large-scale simulations based on Azure Functions traces~\cite{harvested_serverless, serverless_in_the_wild}, where it achieves \textcolor{red}{x\%} and \textcolor{red}{x\%} lower latency than CVMs in cold and warm executions, respectively.
In large-scale simulations using Azure Functions traces~\cite{harvested_serverless, serverless_in_the_wild}, \projectname{} achieves significantly lower invocation latency than CVMs (i.e., 5ms vs. 489s at the 50th percentile,  1.5s vs. 881s at the 99th percentile). 
% \textcolor{red}{x\%} and \textcolor{red}{x\%} lower latency than CVMs in cold and warm executions.
% highlighting its efficiency across all scenarios. 
Further, performance microbenchmarks indicate that \projectname{}'s cold starts are $3.5\times$ faster than CVMs and 35\% than traditional VMs, while warm starts show negligible latency (<10.3 ms).
Function communication latency decreases by $1.4-27\times$ thanks to \projectname{}'s data-centric networking architecture, while function chaining performance improves by $16.7-30.2\times$ over CVMs. 
Attestation latency is also reduced as \projectname{} minimizes the measurement time (0.25 ms in warm starts).
% from 1.1 s to 0.25 ms in warm start cases. %, making verification practical for serverless functions.
Lastly, \projectname{} has a smaller memory footprint than CVMs (up to $907\times$) for the same number of concurrent functions, rendering \projectname{} an ideal solution for secure serverless deployments in untrusted clouds.

\section{Confidential Virtual Machines (CVMs)}
\label{sec:background}

As an alternative to existing cloud serverless virtualization architectures, CVMs~\cite{gcp-confidential-computing, azure-confidential-computing, ibm-confidential-computing} (e.g., AMD SEV-SNP~\cite{sev-snp}, Intel TDX~\cite{tdx}, ARM CCA~\cite{cca}) provide a promising way to address escalating security requirements while preserving compatibility~\cite{misono2024confidential}. 
%Precisely, CVMs protect the VM code, data, and memory from unauthorized access, including hypervisors/VM monitors (VMMs). CVMs use hardware extensions (e.g., AMD SEV-SNP~\cite{sev-snp}, Intel TDX~\cite{tdx}, ARM CCA~\cite{cca}) to ensure data confidentiality and integrity via hardware-enforced isolation and encryption. % and secure interrupt handling.
Precisely, CVMs protect the VM code, data, and memory from unauthorized access, including the hypervisor, via hardware-enforced isolation and encryption.

During a CVM launch, the VMM loads initial guest code, data, and state but cannot access the guest memory or modify its state, beyond this point. 
Guest firmware configures the private memory and performs necessary measurements. The guest kernel must be CVM-aware to interact with the trusted hardware. Host communication uses controlled, encrypted hypercalls.
%At runtime, the guest validates untrusted VMM inputs (e.g., device I/O, ACPI, \texttt{cpuid}, \texttt{rdmsr})~\cite{tdx_guest_hardening}.
%
CVMs support remote attestation to verify their initial state and platform integrity. While providing strong internal security, they require secure protocols (e.g., networking) for data in transit/at rest to maintain their security guarantees.

\revision{
%Further, CVMs support running an enlightened guest OS.
Currently, Linux supports Intel TDX, AMD SEV-SNP, and ARM CCA (the hardware support of ARM CCA is absent as of now).
In addition, Gramine-TDX~\cite{kuvaiskii2024gramine} minimizes the guest TCB by employing a LibOS architecture and specializing the OS to a specific application on the Intel TDX platform.
}

%% old text.
\if 0
Confidential Virtual Machines (CVMs) are hardware-based solutions that protect the code, data, and memory of VMs from unauthorized access, including from hypervisors or VM monitors (VMMs). %(\autoref{fig:cvm-partitioning}). 
% \hyperref[fig:cvm-partitioning]{Figure 1a}). 
CVMs leverage hardware extensions (e.g., AMD SEV-SNP~\cite{sev-snp}, Intel TDX~\cite{tdx}, ARM CCA~\cite{cca}) to ensure data confidentiality and integrity of VMs. Such protections are implemented through hardware-enforced memory and register state encryption, and secure interrupt handling.

The VMM loads the initial guest code, data, and state during launch. 
Beyond this point, the VMM cannot access the guest memory or modify its state. 
The guest firmware configures the CVM-private memory and performs necessary measurements.
The guest kernel must be CVM-aware to interact with the trusted hardware. Communication with the host occurs only through controlled hypercalls via encrypted buffers. 
During runtime, the guest is responsible for validating the untrusted inputs from the VMM (e.g., device I/O, ACPI tables, \texttt{cpuid} and \texttt{rdmsr} instructions)~\cite{tdx_guest_hardening}.

To establish trust, CVMs allow for remote attestation that enables the verification of the CVM's initial state and the integrity of the underlying platform from an external party. While CVMs ensure robust internal security, they must employ secure protocols for data exchanges (e.g., networking) to preserve their guarantees for data in transit and/or at rest. 
\fi

\myparagraph{CVM networking architecture}
%\pramod{moved from s3.2 -- elaborate a bit more for a smooth start -- not too much..}
Since the hypervisor cannot directly access the CVM's memory, CVMs use an unencrypted shared buffer (\emph{bounce buffer}~\cite{bounce_buffer}) for network communication and employ encrypted protocols (e.g., TLS).
Specifically, typical CVM networking involves~\cite{li_bifrost_2023}: \emph{(i)} guest kernel encrypts and copies data to a bounce buffer, \emph{(ii)} hypervisor transfers it to the destination CVM's bounce buffer, and \emph{(iii)} the destination CVM copies and decrypts the data for future consumption. This pipeline causes delays due to multiple copies, cryptographic operations, and heavy context switches~\cite{weisse_hotcalls_2017,misono2024confidential}. %metadata management.

\myparagraph{CVM partitioning}
To improve intra-CVM isolation, CVMs offer a layered privilege model~\cite{sev-snp-vmpl, td-partitioning, cca-plane} beyond ring protections. 
This partitioning assigns each vCPU a privilege level defining its memory access rights and allowed instructions.
The highest privilege level can intercept events from lower levels, enabling the provision of secure services (e.g., vTPM~\cite{vtpm-td, narayan_vtpm_2023}) without hypervisor dependence, and potentially shrinking the TCB by trusting only the most privileged components.
%The highest CVM privilege level can perform operations (e.g., interrupt management, system calls) and offer services (e.g., vTPM) typically provided by the VMM, reducing hypervisor dependence and potentially shrinking the TCB by trusting only the highest privilege level components.

%AMD SEV-SNP and Intel TDX implement their privilege-level hierarchies.
Specifically, AMD SEV-SNP's Virtual Machine Privilege Levels (VMPL)~\cite{snp-white-paper} offer four levels (VMPL-0 to VMPL-3), with VMPL-0 being the most privileged.
Higher levels can modify lower-level permissions. %using \texttt{RMPADJUST}. 
%VMPL-0 software handles critical tasks (e.g., APIC emulation~\cite{sev-es}).
%Permissions are managed via the Rapid Memory Protection (RMP) table~\cite{sev-snp-abi}.
Intel's TD-partitioning~\cite{td-partitioning} enables complete nested virtualization within a CVM, allowing up to three nested VMs with an L1 VMM fully controlling the guests.%, unlike traditional nested virtualization requiring host VMM support~\cite{ben-yehuda_turtles_2010}.

\section{Motivation}
\label{sec:motivation}

\begin{table}[t]
\centering
% \scriptsize
% \footnotesize
\tabfontsize
\caption{TCB size comparison. Presented values are KLoC.}
% \vspace{-.2cm}
%\begin{tabular}{| >{\hspace{-3.5pt}} l <{\hspace{-3.5pt}} || >{\hspace{-3.5pt}} c <{\hspace{-3.5pt}} | >{\hspace{-3.5pt}} c <{\hspace{-3.5pt}} | >{\hspace{-3.5pt}} c <{\hspace{-3.5pt}} | >{\hspace{-3.5pt}} c <{\hspace{-3.5pt}} | >{\hspace{-3.5pt}} c <{\hspace{-3.5pt}} | >{\hspace{-3.5pt}} c <{\hspace{-3.5pt}} |}
\begin{tabular}{>{\hspace{-3.5pt}} l <{\hspace{-3.5pt}} | >{\hspace{-3.5pt}} c <{\hspace{-3.5pt}} | >{\hspace{-3.5pt}} c <{\hspace{-3.5pt}} | >{\hspace{-3.5pt}} c <{\hspace{-3.5pt}} | >{\hspace{-3.5pt}} c <{\hspace{-3.5pt}} | >{\hspace{-3.5pt}} c <{\hspace{-3.5pt}} | >{\hspace{-3.5pt}} c <{\hspace{-3.5pt}} }
 \hline
 \textbf{Baseline} & \makecell{\textbf{Host}\\\textbf{Kernel}}& \makecell{\textbf{Firm-}\\\textbf{ware}} & \makecell{\textbf{Guest}\\\textbf{Kernel}} & \textbf{Runtime} & \makecell{\textbf{Hyper-}\\\textbf{visor}} & \textbf{Total}\\
  \hline
  \hline
  \makecell{LibOS\\(Gramine)} & 1,115 & 903 & --- & 36 & --- & 2,054\\
  \hline
  \makecell{Containers\\(Kata)} & 1,115 & 903 & 1,809 & 8,001 & 1,757 & 13,585\\
  \hline
   \makecell{VM\\(KVM-Linux)} & 1,115 & 903 & 3,177 & 744 & 1,757 & 7,696\\
 \hline
   \makecell{CVM\\(SEV-SNP)} & --- & 903 & 3,177 & 744 & --- & 4,824\\
 \hline 
   \makecell{\projectname} & --- & 332 & --- & 780 & --- & 1,112\\
 \hline 
\end{tabular}
\label{tab:comparison_tcb}
% \vspace{-2mm}
\end{table}

\begin{figure*}[t]
    \centering
    % \begin{subfigure}{0.38\textwidth}
    %     \centering
    %     \raisebox{2.7cm}{
    %     \tabfontsize
    %     \begin{tabular}{| >{\hspace{-3.5pt}} l <{\hspace{-3.5pt}} || >{\hspace{-3.5pt}} c <{\hspace{-3.5pt}} | >{\hspace{-3.5pt}} c <{\hspace{-3.5pt}} | >{\hspace{-3.5pt}} c <{\hspace{-3.5pt}} | >{\hspace{-3.5pt}} c <{\hspace{-3.5pt}} | >{\hspace{-3.5pt}} c <{\hspace{-3.5pt}} |}
    %      \hline
    %      \textbf{Baseline} & \makecell{\textbf{Host}\\\textbf{Kernel}}& \makecell{\textbf{Firm-}\\\textbf{ware}} & \makecell{\textbf{Guest}\\\textbf{Kernel}} & \makecell{\textbf{Run-}\\\textbf{time}} & \makecell{\textbf{Hyper-}\\\textbf{visor}}\\
    %       \hline
    %       \makecell{LibOS\\(Gramine)} & 1,115 & 903 & --- & 36 & ---\\
    %       \hline
    %       \makecell{Containers\\(Kata)} & 1,115 & 903 & 1,809 & 8,001 & 1,757\\
    %       \hline
    %        \makecell{VM\\(KVM-Linux)} & 1,115 & 903 & 3,177 & 744 & 1,757\\
    %      \hline
    %        \makecell{CVM\\(SEV-SNP)} & --- & 903 & 3,177 & 744 & ---\\
    %      \hline 
    %        \makecell{\projectname} & --- & 332 & --- & 780 & ---\\
    %      \hline 
    %     \end{tabular}
    %     }
    %     \vspace{-0.6cm}
    %     \caption{TCB size comparison (KLoC).}
    %     \label{tab:comparison_tcb}
    % \end{subfigure}
    \hfill
    \begin{subfigure}{0.3\textwidth}
        \centering
        \includegraphics[width=\textwidth]{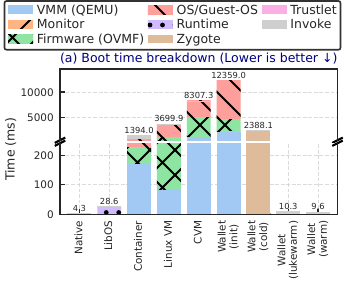}
        \vspace{-0.6cm}
        \phantomcaption
        %\caption{Boot time.}
        % \caption{}
        \label{fig:mot:boot}
    \end{subfigure}
    \hfill
    \begin{subfigure}{0.3\textwidth}
        \centering
        \includegraphics[width=\textwidth]{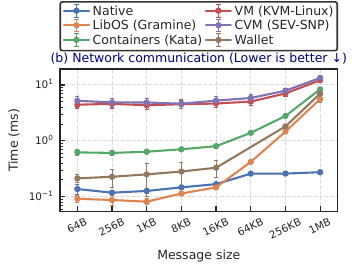}
        \vspace{-0.6cm}
        %\caption{Network communication overheads.}
        % \caption{}
        \phantomcaption
        \label{fig:mot:ipc}
    \end{subfigure}
       \hfill
    \begin{subfigure}{0.3\textwidth}
        \centering
       \includegraphics[width=\textwidth]{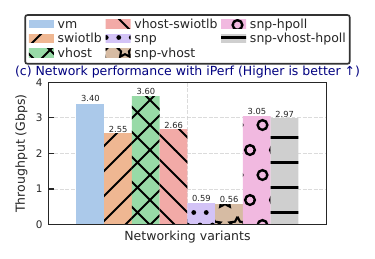}
        %\vspace{-0.6cm}
        \vspace{0.05cm} % add a space as this plot has no xlabel
        \phantomcaption
        % \caption{}
        \label{fig:mot:cvm_network}
    \end{subfigure}
    % \vspace{-0.1cm}
    \caption{Serverless requirement analysis: \emph{(a)}~boot time analysis, \emph{(b)} networking overheads and \emph{(c)}~CVM networking analysis.
    %\masa{the caption duplicates the subcaption. somehow change this} \pramod{or simply remove the second-level caption.}
    }
    \label{fig:combined}
    % \vspace{-3mm}
\end{figure*}

%Confidential computing~\cite{confidential_computing} offers a promising way to protect serverless workloads in the untrusted cloud. Encapsulating the serverless stack (i.e., runtime, functions) within CVMs provides isolation and compatibility with virtualization infrastructures. 

% Confidential computing~\cite{confidential_computing} offers a promising way to secure serverless workloads in the cloud by isolating the serverless stack within CVMs while ensuring compatibility with virtualization.
% However, CVM-based confidential serverless computing faces \emph{six} key challenges due to mismatches between CVM constraints and serverless demands, which we analyze next (experimental setup is described in \autoref{sec:evaluation}).

Confidential computing~\cite{confidential_computing} offers a promising way to secure serverless workloads in the untrusted cloud. 
However, CVM-based serverless computing faces \emph{six} key challenges due to mismatches between CVM constraints and serverless requirements, analyzed below (experimental setup is described in \autoref{sec:evaluation}).

%CVM-based confidential serverless computing faces challenges for sensitive, low-latency, and scalable workloads~\cite{serverless_in_the_wild}. We  experimentally analyze (setup is described in \autoref{sec:evaluation}) to highlight six key mismatches with serverless demands.

%owever, it is not ideal for sensitive serverless workloads requiring strong security, low latency, and scalability~\cite{serverless_in_the_wild}. Below, we experimentally analyze (setup is described in \autoref{sec:evaluation}) and discuss six key issues with CVM-based confidential serverless computing due to mismatches between CVM constraints and serverless demands.

\subsection{Issue \#1: Large Trusted Computing Base}
\label{subsec:mot:tcb}
Minimizing the TCB is crucial for reducing the attack surface in confidential serverless computing. 
% While CVMs remove hypervisor dependencies, CVM-hosted serverless stacks still include in their large TCB monolithic legacy software (i.e., guest OS, firmware) with redundant drivers and services for backward compatibility~\cite{kuvaiskii2024gramine}. 
While CVMs eliminate hypervisor dependencies, CVM-hosted serverless stacks retain large TCBs containing monolithic legacy software with redundant components for backward compatibility~\cite{kuvaiskii2024gramine}.

% still rely on  full guest OSes with unnecessary legacy components and redundant services for backward compatibility~\cite{kuvaiskii2024gramine}.
% their reliance on full guest OSes introduces unnecessary dependencies on a large codebase. 

% CVM-hosted serverless stacks confine their TCB to monolithic legacy software (i.e., guest OS, firmware) with redundant drivers and services for backward compatibility~\cite{kuvaiskii2024gramine}. 
\autoref{tab:comparison_tcb} showcases that the CVM TCB size (4.8M LoC) is smaller than traditional VMs (7.7M LoC) but much larger than LibOS alternatives (e.g., Gramine LibOS: 2M LoC).
Thus, CVM-hosted serverless stacks do not adequately minimize the TCB. 
The inclusion of a full guest OS increases complexity and exposes a large attack surface.
In contrast, \projectname{} achieves approximately 4$\times$ TCB reduction compared to CVMs and 2$\times$ compared to Gramine LibOS without compromising security.

\subsection{Issue \#2: Slow Boot Times}
\label{subsec:mot:boot_times}

Low startup latency is vital for serverless functions, but CVMs introduce inherent boot penalties~\cite{severifast, misono2024confidential} due to three security mechanisms~\cite{sev-snp-abi,tdx-module}:
% CVMs introduce inherent penalties to the boot process~\cite{severifast, misono2024confidential}, which are attributed to three core security mechanisms~\cite{sev-snp-abi,tdx-module}: 
\emph{(i)} costly per-guest key generation and CVM memory pre-encryption, \emph{(ii)} explicit guest state/memory initialization and boot measurement calculation for attestation, and \emph{(iii)} synchronous signed report generation and external validation delaying function execution.

To this end, we measure the boot time of different virtualization architectures (\autoref{fig:mot:boot}), which indicate that CVMs exhibit a boot time of $\sim$8.3 seconds, 2.2$\times$ slower than standard VMs (3.7s), 
primarily due to VMM setup, OS initialization, and encrypted memory configuration.
% \revision{The CVM boot incurs delays for three core security mechanisms~\cite{sev-snp-abi,tdx-module}: 
% \emph{(i)} the trusted component, such as the AMD Platform Security Processor (ASP) or TDX module, must generate per-guest keys and pre-encrypt the CVM memory pages, 
% \emph{(ii)} the hypervisor must explicitly initialize the guest state and its memory, which involves the calculation of their boot measurements for later attestation,
% and \emph{(iii)} the attestation process mandates synchronous signed report generation by the trusted component and external validation, which can further block function execution until completion.}
% CVMs exhibit a start time of $\sim$8.3 seconds, 2.2$\times$ slower than standard VMs (3.7 s).
% This delay primarily stems from VMM setup, guest OS initialization, and encrypted memory configuration.
These inevitable security-induced boot delays often exceed function lifetimes, making CVMs impractical for latency-sensitive serverless computing.
% Therefore, CVMs' strong security causes inevitable boot latency costs, potentially exceeding function lifetimes, making them impractical for latency-sensitive serverless computing. 
\revision{While \projectname{} has overall the highest initialization time ($\sim$~12s), it is an one-time cost.}
In contrast, with an initialized \projectname{}, we can achieve faster cold starts (2.4 s) than traditional VMs (3.7 s), and a near-negligible warm start latency (<10.3 ms). 
\revision{Notably, this warm start achieves even lower start-up times compared to the LibOS ($\sim$ 28 ms) on its own running without isolation on the host, thanks to its efficient CoW-based function execution. }

\subsection{Issue \#3: High Networking Overheads}
\label{subsec:mot:network_communication}

High-performance networking is crucial in distributed serverless applications with frequent data exchanges (e.g., function chaining). However, CVMs introduce significant networking overheads due to strict isolation and the requirement for additional encryption (e.g., TLS) to secure data in transit, which involves heavy context switches~\cite{cheng_tdxsurvey_2023,misono2024confidential}.

%\pramod{can be moved into the background, under CVM networking}

%To gauge these overheads, we present the communication costs between two separate function instances using the same virtualization model and varying message sizes in \autoref{fig:mot:ipc}. 
To gauge these overheads, we measure the communication costs between two instances of the same virtualization model with varying message sizes (\autoref{fig:mot:ipc}). 
We observe that CVMs consistently exhibit higher latency compared to traditional VMs (9-20\%) and lightweight solutions such as Kata containers and Gramine (up to 59$\times$). 

%\pramod{Add a new plot for the CVM netowork analysis and then expalin the CVM networking bottleneck analysis.}
We further examine the network performance of CVMs with various Linux configurations.
\autoref{fig:mot:cvm_network} shows iPerf~\cite{iperf} throughput (UDP, 1460 bytes payload, client runs on the host). 
``swotlb'' means that bounce buffer is enabled, and ``vhost'' implies using the vhost optimization.
The SNP VM \textit{always} uses a bounce buffer.
We observe that the SNP VM suffers from performance degradation (only 17\% of the VM), and the guest-side halt polling~\cite{guest-halt-polling} (``-hpoll'') mitigates the performance overhead at the cost of excessive CPU usage~\cite{misono2024confidential}.
% XXX: explain/justify why show UDP measurement here
%
%Current CVM stacks transpose VM-centric protocols to serverless contexts, treating each function as a tenant in a different CVM.
%This enforces unnecessary data encryption even for co-located functions, hurting performance.
%In serverless environments,  where chaining multiple functions is common, the CVM communication overhead can become a significant bottleneck.
In contrast, 
\projectname{} achieves lower latency than CVMs (1.4-27$\times$) leveraging function co-location and following a data-centric networking model for secure, yet performant, communication.

\begin{figure*}[t]
\centering
\begin{subfigure}{0.33\textwidth}
    \centering
    % 5 nodes, 500 execution slots, 500 cache size, 10 mins cache time
    \includegraphics[width=\textwidth]{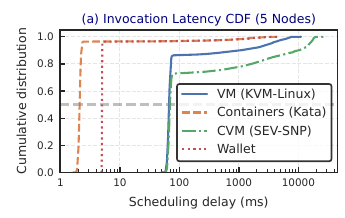}
    \vspace{-0.7cm}
    % \caption{Scheduling delay.}
    % \caption{}
    \phantomcaption
    \label{fig:mot:scheduling_delay_cdf}
  %  \Description[<short description>]{<long description>}
\end{subfigure}
\hfill
\begin{subfigure}{0.33\textwidth}
    \centering
    \includegraphics[width=\textwidth]{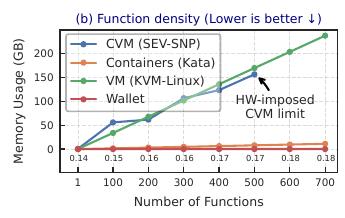}
    \vspace{-0.7cm}
    % \caption{Function density.}
    % \caption{}
    \phantomcaption
    \label{fig:mot:function_density}
  %  \Description[<short description>]{<long description>}
\end{subfigure}
\hfill
\begin{subfigure}{0.33\textwidth}
    \centering
    \includegraphics[width=\textwidth]{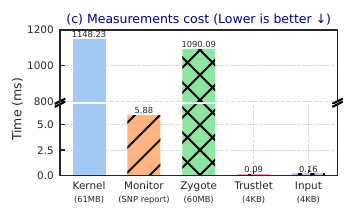}
    \vspace{-0.7cm}
    % \caption{Attestation measurement costs.}
    % \caption{}
    \phantomcaption
    \label{fig:mot:attestation}
   % \Description[<short description>]{<long description>}
\end{subfigure}
% \vspace{-0.4cm}
\caption{Serverless requirement analysis: \emph{(a)}~the scheduling delay, \emph{(b)}~the function density, and \emph{(c)}~attestation measurements.}
% \vspace{-6mm}
\end{figure*}

\subsection{Issue \#4: Inefficient Scheduling}
\label{subsec:mot:scheduling}
Efficient scheduling is a critical challenge~\cite{dirigent, serverless_scheduler_costs, jiagu} in confidential serverless computing. 
While optimized two-level (global/local) schedulers~\cite{faasm,pheromone,mesos} favor warm starts, the limited CVMs per node~\cite{amd_sev_manual, sev_asid_github_issue, tme_key_limit, tme_tdx_correlation} can lead to more cold starts, increasing invocation latency. 
Further, traditional local schedulers, such as the Completely Fair Scheduler (CFS), are also ineffective~\cite{serverless_scheduler_costs} for short functions~\cite{ context_switch_cost}.
%(e.g., due to frequent context switches~\cite{ context_switch_cost}).

Specifically, confidential serverless functions in isolated environments cannot share resources, restricting the flexibility of resource-aware schedulers~\cite{faasm}. 
%The limited pool of instances per CVM node forces balancing instance reuse and locality, and increases cold-to-warm start ratios, which further raises invocation latency~\cite{serverless_scheduler_costs,jiagu}. 
The limited pool of instances per CVM node forces balancing instance reuse, increasing cold start ratios~\cite{serverless_scheduler_costs,jiagu}. 
In addition, switching execution context between CVMs incurs significant overheads (e.g., TLB/cache flushing, memory validation)~\cite{cpc,misono2024confidential, verismo}. 
Our large-scale system simulation using sampled~\cite{ustiugov:invitro} Azure Functions traces~\cite{harvested_serverless, serverless_in_the_wild} (30 minutes, 4,000 functions, 4.1 million invocations) shows CVMs having higher scheduling latency across all percentiles compared to VMs and microVMs (\autoref{fig:mot:scheduling_delay_cdf}) under the same scheduling model.

% To analyze this, we conduct a large-scale system simulation using a sampled Azure Functions trace~\cite{harvested_serverless, serverless_in_the_wild} (30 minutes, 4,000 functions, 3.8 million invocations, sampled with the In-Vitro~\cite{ustiugov:invitro}).  
% \autoref{fig:mot:scheduling_delay_cdf} shows that CVMs have a higher scheduling latency across all percentiles compared to VMs and microVMs under the same workload and scheduling model.
% 5 nodes, 500 execution slots, 500 cache size, 10 mins cache time

Thus, we conclude that standard scheduling is suboptimal for CVM-based serverless. 
\projectname{} achieves lower scheduling latency with better tail latencies by sustaining more warm instances per node and using a run-to-completion model~\cite{serverless_scheduler_costs}.

\subsection{Issue \#5: Impractical Consolidation}
\label{subsec:mot:server_consolidatiion}
Efficiently packing multiple functions onto shared hardware resources is vital in serverless computing~\cite{firecracker, rund}, but also challenging for monolithic CVM-based serverless deployments for three fundamental reasons.
%
%First, CVMs lack effective memory deduplication~\cite{severifast}, wasting memory resources even when multiple instances of similar functions are deployed. 
%Encrypted shared memory is technically possible~\cite{encrypted_dedup, amd_sev_manual, tme_key_limit, tme_tdx_correlation} but would violate the isolation requirements of CVM-hosted serverless deployments. 
First, CVM's encrypted memory makes effective memory deduplication difficult, if not technically impossible~\cite{encrypted_dedup, amd_sev_manual, tme_key_limit, tme_tdx_correlation}, as each memory is encrypted with a different key.
Second, hardware limits the number of CVMs per node~\cite{amd_sev_manual, sev_asid_github_issue, tme_key_limit, tme_tdx_correlation}, creating a hard ceiling on function density. 
Lastly, CVMs' heavy-weight stacks cause higher per-function memory overhead, reducing achievable density considering a fixed hardware budget.

\autoref{fig:mot:function_density} presents the memory usage of concurrent functions in a node. CVMs have the steepest increase, consuming $\sim${168} GB for {500} functions. The respective number for MicroVMs is $\sim${8.5} GB. 
% CVMs consume {\textcolor{red}x-x$\times$} more memory than VMs due to their inability to deduplicate encrypted memory. 
\autoref{fig:mot:function_density} further highlights the limit of CVMs per node imposed by the hardware ($\sim$500 in our case).

%Thus, CVM-based serverless struggles with function consolidation due to resource management inefficiency and hardware constraints.
Thus, CVM-based serverless struggles to consolidate functions efficiently. %consolidation due to resource management inefficiency.
In contrast, \projectname{} can host multiple isolated functions within a CVM (up to 907$\times$ less memory than CVMs), exceeding the hardware-imposed density limit.

\subsection{Issue \#6: High Attestation Overheads}
\label{subsec:mot:attestation}
%Remote attestation~\cite{menetrey_attestation_2022} enables trust establishment with a CVM, but its second-scale latency per CVM instance is excessive for ms-scale serverless functions.
Remote attestation~\cite{menetrey_attestation_2022} establishes trust in a CVM, but its second-scale latency per CVM is excessive for ms-scale serverless functions.
Specifically, attestation involves measuring software components and generating a signed report.
To ensure function execution integrity, additional measurements of the function and input/output are necessary.
This process can become costly, especially in function chaining scenarios.
%CVM attestation involves a trusted hardware component and software measuring the runtime components.
%The trusted component generates a signed report with firmware, memory, and initial state measurements for external verification. 
%This process focuses on boot-time verification and requires a complete re-attestation per function invocation, which becomes costly, especially in function chaining scenarios.
% Function chaining further complicates where multiple attestations must be performed sequentially or concurrently.

\autoref{fig:mot:attestation} shows that the measurement of a full guest OS kernel for a Linux CVM can take up to 1.1 s, significantly increasing the overall startup latency. 
While long-running workloads can amortize this cost, the short execution times of serverless functions make the repetitive attestation overhead prohibitive. 
Our \projectname{} prototype significantly reduces the measurement cost to $\sim$0.25 ms for warm starts by reusing already calculated measurements across the same functions.

\subsection{Summary and Problem Statement}
CVMs face critical limitations when employed in serverless settings. Their large TCB, slow boot times, I/O communication overhead, and costly attestation render them ill-suited for serverless computing. These constraints further lead to inefficient scheduling and low function density due to hardware limitations and suboptimal resource management.

\if 0
\revision{
To this end, there exists prior work that focuses on providing specialized solutions to tackle these challenges.
Specifically, Gramine-TDX~\cite{kuvaiskii2024gramine} minimizes the TCB, whereas SEVeriFast~\cite{severifast} optimizes CVM startup time.
However, none of them addresses the presented challenges specific to serverless computing in their entirety. % of network overheads, inefficient scheduling, impractical consolidation, and high attestation overheads.
}
\fi

\revision{
To this end, there exists prior work that focuses on providing specialized solutions to tackle these challenges~\cite{kuvaiskii2024gramine, severifast}.
Specifically, Gramine-TDX~\cite{kuvaiskii2024gramine} provides a slim, security-first OS kernel, based on Gramine LibOS~\cite{tsai_gramine_2014}, to run unmodified applications in CVMs while having a minimal TCB.
However, it is mainly suited for single-tenant architectures, where each serverless function requires its dedicated CVM, which conflicts with confidential serverless computing requirements. 
}

\revision{
Precisely, Gramine-TDX cannot achieve efficient function consolidation due to hardware CVM limits, requires per-function attestation, and forces inter-function communication through expensive CVM networking. Further, Gramine-TDX cannot leverage function co-location opportunities within the same CVM in a secure manner, as it operates with system-wide privileges and lacks the fine-grained privilege separation mechanisms necessary to safely isolate multiple co-located functions from each other and from the underlying software layers. These architectural constraints make Gramine-TDX unsuitable for the multi-tenant, high-density, and communication-intensive nature of serverless workloads.
}

\if 0
To this end, several prior studies~\cite{severifast,kuvaiskii2024gramine} provide specialized solutions to tackle these challenges. However, none fully addresses the presented challenges specific to serverless computing in their entirety.
Specifically, SeVeriFast~\cite{severifast} focuses on optimizing the SEV-SNP VM startup time. 
On the other hand, Gramine-TDX~\cite{kuvaiskii2024gramine} adopts a LibOS architecture, specializing the guest OS for a specific application to minimize its TCB.
While these efforts partly address some of the aforementioned challenges in serverless computing, they are limited to supporting only running single-function execution in a single CVM instance. Therefore, the scalability issue still remains, along with  high networking and attestation overhead.
\fi

% from related work:
%Unlike Gramine-TDX, \projectname{}'s LibOS is only running in a certain privilege level within the CVM and does not have system-wide privileges. Further, in contrast to Gramine with its existing backends, \projectname{} provides a framework for managing multiple instances within the same CVM, enabling multitenancy and flexible interconnections among trusted processes while maintaining instance protection.

%\boundedbox{
\myparagraph{Problem statement}
To materialize confidential serverless computing, we have to answer a critical question: \emph{how can we execute lightweight, short-lived functions in the untrusted cloud in a secure and verifiable manner while complying with the performance and scalability requirements?} 
%}

%% old text -- summary
\if 0
\subsection{Summary and Problem Statement}
CVMs face critical limitations when employed in serverless settings. Their large TCB, prohibitively high boot latency, secure communication overheads, and costly attestation processes render them ill-suited for the serverless computing paradigm. 
These constraints further lead to inefficient scheduling and low function density due to hardware limitations and suboptimal resource management.

\boundedbox{
\myparagraph{\underline{Problem statement}}
To materialize confidential serverless computing, we have to answer a critical question: \emph{how can we execute lightweight, short-lived functions in the untrusted cloud in a secure and verifiable manner while complying with the performance and scalability requirements?} 
}

\fi 
\section{Overview}
\label{sec:overview}

% To address confidential serverless computing needs, we propose \projectname{}, a system for untrusted clouds. % for real? 
As a solution, we present \projectname{}, a lightweight confidential computing system for secure serverless deployments in untrusted cloud environments.
%Below, we detail its architecture (\autoref{subsec:architecture}), the lifecycle of a serverless request (\autoref{subsec:lifecycle}), its threat model (\autoref{sec:overview:thread_model}) and highlight its design primitives (\autoref{subsec:design_primitives}).

%% old text -- opening.
%To meet the requirements of confidential serverless computing, we propose \projectname{}, a confidential serverless computing system for untrusted cloud environments. Below, we present the system architecture of \projectname{} (\autoref{subsec:architecture}),  the lifecycle of a serverless request (\autoref{subsec:lifecycle}),  and, lastly, highlight its design primitives (\autoref{subsec:design_primitives}).

\subsection{Threat Model}
\label{sec:overview:thread_model}
\projectname{} extends the CVM threat model~\cite{sev-snp, cheng_tdxsurvey_2023, tdx_google_security_review, linux_coco_threat_model} by excluding the guest OS from its trust boundaries. We consider adversaries controlling the guest OS attempting to compromise serverless functions within trusted processes, the host OS attacking the CVM via VM-VMM interfaces (e.g., network I/O), or deploying malicious functions to access neighboring functions and leak data. 
Users trust function providers, but serverless and platform providers are potentially malicious.
% They might also compromise the host OS to attack the CVM via VM-VMM interfaces (e.g., network I/O) or deploy malicious functions to access neighboring functions and leak user data. While users trust function providers, serverless and platform providers are considered potentially malicious.

The trusted monitor is \projectname{}'s \emph{only} trusted software, enabling remote attestation. We assume that the platform hardware/firmware functions correctly and deployed functions do not intentionally leak data. Users must employ encrypted protocols (e.g., TLS) for communication. %, as \projectname{} does not provide security over the network.
\projectname{} does not address denial-of-service, physical, and side-channel attacks.

% Wallet overview
\begin{figure}[t]
     \centering
     \includegraphics[width=\columnwidth]{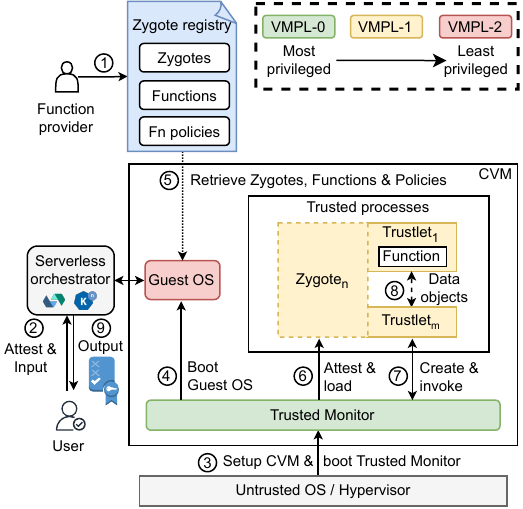}
     % \vspace{-2mm}
     \caption{\projectname system architecture overview. 
     %{\em The {trusted monitor} securely deploys and isolates serverless functions, the {trusted processes} enable function execution within an isolated environment, the untrusted {guest OS} hosts the serverless management system, and the {zygote registry} provides template images for serverless functions.}
     }
     % \pramod{ToDo} {\em explain the high level components.}}
     \label{fig:overview:overview}
    % \Description[<short description>]{<long description>}
     % \vspace{-3mm}
\end{figure}

\subsection{System Architecture}
\label{subsec:architecture}
\projectname{} builds on CVMs and uses hardware-enforced privilege partitioning (e.g., VMPL~\cite{snp-white-paper}, TD-partitioning~\cite{td-partitioning}) to create lightweight \emph{trustlets} --- minimal serverless processes that execute within a secure environment inside a CVM. 
%secure execution environments tailored for serverless functions. 
% \pramod{we should tie it to the abstraction of trusted, both here and in the intro, as mentioned in the paper title.} 
\autoref{fig:overview:overview} presents \projectname{}'s key components: the trusted monitor, the trusted processes, the data objects and the zygote registry.

The \textbf{trusted monitor} is a small, privileged, and nested software component within \projectname{}'s TCB. 
% that operates at the highest privilege level. 
It manages the instantiation and invocation of serverless functions within lightweight protection domains isolated from the guest OS. It performs security-critical tasks such as memory management, function deployment, and attestation. % using hardware-enforced privilege partitioning.

Within the CVM, \projectname{} introduces the notion of \textbf{trusted processes} for hosting serverless functions. These processes possess enclave-like features: \emph{(i)} isolation from the guest OS, \emph{(ii)} trusted boot, and \emph{(iii)} remote attestation capabilities. 
They are categorized into \emph{zygotes} and \emph{trustlets}. 
Zygotes serve as pre-initialized templates containing runtime environments (e.g., Python), while trustlets are lightweight instances, derived from zygotes, to execute functions in an isolated manner. 
To facilitate inter-function communication, \projectname{} provides the abstraction of \textbf{data objects}, managed by the trusted monitor ensuring isolation.
This design favors reduced boot times, efficient memory use via copy-on-write~\cite{sand,catalyzer,sock,du_xpusim_2022,li_pagurus_2022}, and optimized data exchange between co-located functions.

Meanwhile, the untrusted guest OS runs at a lower privilege level and mediates communication between the serverless orchestration framework and the trusted monitor. It lies out of the TCB and is responsible for retrieving zygote images and functions from the \textbf{zygote registry} and providing them to the monitor for validation, loading, and execution.

\begin{figure}[t]
     \centering
     \includegraphics[width=\columnwidth]{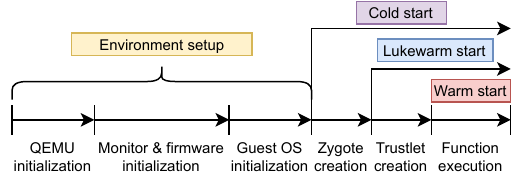}
     % \vspace{-2mm}
     \caption{\projectname{} invocation phases.
     %\textit{\projectname{} optimizes long environment setup and cold-start latency by employing fork-based Trustlet creation (Lukewarm start).}
     }
     \label{fig:invocation_phases}
     % \vspace{-3mm}
\end{figure}

\subsection{Life of a Request}
\label{subsec:lifecycle}
%\autoref{fig:overview:overview} illustrates how \projectname{} handles function deployment and invocation.
\autoref{fig:overview:overview} illustrates the life of a request in \projectname{}.
First, the serverless provider publishes a zygote image with the function runtime in the zygote registry. Function providers develop functions and associate them with a zygote (\Circled{1}), and generate a function key pair for user request encryption.

\myparagraph{Initial environment setup}
For a function request, users obtain the function provider's public key, encrypt their request where they include a symmetric key for the encryption of the result, and submit it to the serverless orchestrator (e.g., OpenWhisk~\cite{openwhisk}, Knative~\cite{knative}) (\Circled{2}). 
\revision{
%This happens with the assumption that the user trusts the function provider.
This allows the function to directly encrypt with the key of the function provider. 
%Alternatively, if no initial trust is given to the function provider, a test function call could be made, which would return with the attestation report, allowing the user to verify that the correct function is  called.
}
The orchestrator checks the available \projectname{} instance, and if none, launches a CVM with the trusted monitor (\Circled{3}). 
%it launches a new \projectname{} instance with the trusted monitor (step \Circled{3}). 
%If no instance exists, it launches a CVM with the trusted monitor as its boot image (step \Circled{3}). 
The trusted monitor then loads the guest OS, where the serverless management system runs, into a lower privilege level (\Circled{4}).
\revision{\projectname{} instances can be created proactively by the serverless provider, and its expensive initialization step occurs only once.}

\myparagraph{Handling function request}
Once initialized, the guest OS handles the incoming request, forwarded by the monitor. 
\projectname{} has three types of request invocation: \emph{cold}, \emph{lukewarm} and \emph{warm}  (\autoref{fig:invocation_phases}).
\revision{The cold start breakoff point in \projectname{}'s workflow is right after the guest OS initialization. The trusted monitor and guest OS constitute the environment for a function execution runtime in the same way a host OS is the environment for the Docker runtime.}
A cold start occurs when the required zygote is not loaded into the trusted monitor. In this case, the trusted monitor fetches the zygote and function from the registry (\Circled{5}).
Then, the function provider attests \sysname, establishes secure communication with the trusted monitor, and shares their secrets, including the function's private key and measurements of the zygote image and the function.
The guest OS then instructs the trusted monitor to load the zygote (\Circled{6}), which is performed after verifying the zygote's integrity against the provided measurement.

At this point, \projectname{} can process requests. 
If the function's zygote exists (\emph{lukewarm start}), \projectname{} spawns a trustlet for function execution (\Circled{7}). 
Subsequent invocations within a running trustlet (\emph{warm start}) execute without setup delays. 
% The guest OS forwards user data to the trusted monitor, which decrypts it and invokes the trustlet. 
The guest OS forwards user data to the trusted monitor, which creates input data objects, decrypts the data, and invokes the trustlet.
% During execution, the trusted monitor can create secure data objects to facilitate low-latency communication between co-located trustlets, enabling efficient function chaining (\Circled{8}).
During execution, the trustlet reads from the input and creates output data objects, which can act as input to other functions, thus enabling efficient function chaining (\Circled{8}).
Importantly, the trusted monitor mediates this access to ensure isolation between co-located trustlets. 
Once execution completes, the monitor encrypts the result from output data objects with the request's symmetric key and returns it alongside a signed attestation report to the user via the guest OS (\Circled{9}).

\subsection{Design Principles and Primitives}
\label{subsec:design_primitives}

% Having detailed \projectname{}'s architecture and workflow, 
We now present the design principles and primitives of \projectname{} that address the six key problems, outlined in \autoref{sec:motivation}. %of CVM-based serverless computing

\myparagraph{\#1: Nested confidential execution} As a countermeasure to the large TCB, the hardware-limited scalability and the costly inter-function communication of CVM-based serverless deployments, \projectname{} introduces nested confidential execution. 
Inspired by nested virtualization~\cite{ben-yehuda_turtles_2010, nested_vmx, nested_google, nested_azure, zhang_cloudvisor_2011}---not directly applicable to CVMs due to host VMM support requirements~\cite{vtz, veil_ahmad_2024}---\projectname{} leverages intra-CVM isolation~\cite{sev-snp-vmpl, td-partitioning, cca-plane} to partition runtime components~\cite{cabin}. % into privilege domains~\cite{cabin}. 
This allows the secure coexistence of untrusted components in lower-privilege compartments, while critical services % (e.g., interrupt handling) 
operate in highly privileged domains, reducing the software TCB.

In this model, the \emph{trusted monitor} governs the nested execution environments and operates at the highest privilege level, managing security-critical tasks while restricting untrusted components to lower levels via strict, hardware-enforced page-level access control~\cite{sev-snp-vmpl, td-partitioning}. Further, it optimizes inter-function communication with data exchange via shared memory, realizing fast chaining for co-located functions.

Nested confidential execution also improves scalability without compromising security. It enables higher function density within a single CVM compared to conventional deployments, which are constrained by hardware limits on concurrent CVM instances per node~\cite{amd_sev_manual, sev_asid_github_issue, tme_key_limit, tme_tdx_correlation}.

\myparagraph{\#2: Decoupled guest OS architecture} 
Conventional CVMs include the, often bloated~\cite{severifast}, guest OS in their TCB. %, causing prolonged boot times (i.e., full OS initialization).
\projectname{} decouples the \emph{guest OS} from trusted operations and removes its boot and measurement process from the function invocation path, mirroring %, even for cold starts. 
% This compartmentalization mirrors 
ARM TrustZone's principles, i.e., secure/normal world separation~\cite{trustzone, trustzone_explained},  as performed in prior works, such as TLR (Trusted Language Runtime)~\cite{tlr}.
% \dimitris{add TLR spelled out}

After trusted monitor initialization, \projectname{} boots the untrusted guest OS, totally isolated from function execution environments.
%The trusted monitor handles function scheduling and memory allocation without guest OS intervention. % excluding it from the TCB and reducing the attack surface.
% Precisely, after the trusted monitor initialization, a \projectname{} instance boots the untrusted guest OS. 
% The trusted monitor enforces strict isolation between the guest OS and users' function execution environments and handles function scheduling and memory allocation, without guest OS intervention.
This decoupling excludes the guest OS kernel from the TCB, significantly reducing the attack surface and ensuring the guest OS vulnerabilities do not compromise the security of serverless functions.
The guest OS handles \emph{only} auxiliary tasks such as receiving requests and forwarding function invocations to the trusted monitor, maintaining compatibility with existing serverless frameworks. This reduces startup times compared to CVM-based deployments where cold starts involve full guest OS initialization and measurement.

\myparagraph{\#3: Trusted process templates} 
Existing confidential serverless deployments must initialize each function instance from scratch. 
Startup optimizations (e.g., snapshotting) are impractical~\cite{severifast} due to memory confidentiality (e.g., encryption) requirements.
% leading to high startup and resource allocation overhead. 
To this end, \projectname{} introduces \emph{trusted processes} as the core abstraction for executing serverless functions.

These processes—categorized as \emph{zygotes} and \emph{trustlets}, inspired by Android's process model~\cite{zygote, lee_morula_2014}—operate at a dedicated privilege level within isolated address spaces.
Zygotes serve as pre-initialized templates with runtime dependencies, while trustlets instantiate from zygotes using copy-on-write mechanisms~\cite{sand,catalyzer,sock,du_xpusim_2022,li_pagurus_2022}, enabling concurrent function launching with efficient memory usage and high density.
% Trusted processes are categorized into \emph{zygotes} and \emph{trustlets}, inspired by Android's zygote process~\cite{zygote, lee_morula_2014} for efficient process spawning. 
%In \projectname{},
% They operate at a dedicated privilege level %(higher than the guest OS, lower than the trusted monitor)
% within isolated address spaces. % managed page tables configured by the trusted monitor.
%
% Essentially, zygotes are pre-initialized templates containing runtime dependencies (e.g., Python). 
%Trustlets are instantiated from zygotes using copy-on-write mechanisms~\cite{sand,catalyzer,sock,du_xpusim_2022,li_pagurus_2022} and execute user functions in isolation. % environments. 
%This allows for launching multiple functions concurrently in a memory-efficient manner in a single CVM, achieving high density. 
% Trustlets are instantiated from zygotes using copy-on-write mechanisms~\cite{sand,catalyzer,sock,du_xpusim_2022,li_pagurus_2022}, allowing for launching multiple functions concurrently in a memory-efficient manner in a single CVM, achieving high density. 
% To facilitate efficient function chaining, the trusted monitor provides them with secure \emph{shared memory channels}. % XXX: secure channel is explained in the #5: Hybrid I/O architecture

Building on this execution model, \projectname{} implements its \emph{differential attestation} protocol, which incrementally builds cumulative trust chains by reusing zygote and function measurements, requiring only the measurement of mutable components (e.g., input/output) during invocation, thus reducing attestation latency for both standalone and chained functions.

% Building on this execution model, \projectname{} implements its \emph{differential attestation} protocol that incrementally constructs cumulative trust chains by reusing zygote and function measurements. Thus, on a function invocation, \projectname{} requires solely the measurements of mutable components (e.g., input/output) for its attestation report, reducing the latency both for standalone functions and chained workflows. %, while ensuring end-to-end verifiability.
% \pramod{underselling differential attestation...}

\myparagraph{\#4: Dynamically loadable LibOS architecture} 
To minimize function startup latency and its TCB while supporting diverse workloads (e.g., Python, Node.js), \projectname{} employs a dynamically loadable LibOS architecture~\cite{kylinx, ebbrt}, adopting concepts of unikernel-based systems~\cite{unikernels, ukl, kylinx}. 
However, instead of requiring developers to build custom LibOS images, \projectname{} dynamically loads lightweight LibOS-based runtimes tailored to specific functions at runtime. 
\revision{By providing the language runtimes that directly call a function, common serverless functions can be immediately used inside \projectname{}.}% without major modifications. }

The LibOS forms the core of \projectname{}'s trusted processes, providing them with essential runtime functionalities. By dynamically loading minimal runtimes rather than full OSes or containers, \projectname{} reduces initialization overhead. % and the components to be measured for attestation.
This design allows on-demand instantiation of lightweight, pre-measured templates and avoids redundant software measurement steps during function invocation, further lowering startup latency %while supporting various languages and runtimes
without increasing developer effort.

\myparagraph{\#5: Confidential networking architecture} 
CVM I/O stacks~\cite{virtio_spec, vsock} suffer from performance penalties due to data copies, VM exits, and high CPU utilization for cryptographic operations~\cite{li_bifrost_2023,misono2024confidential, efficient_paravirt}. 
\projectname{}'s key insight is to leverage its function density (\autoref{subsec:mot:server_consolidatiion}) and provide a monitor-mediated, high-performance networking architecture for CVMs. 

\projectname{} leverages \textit{function co-location} and employs a \emph{data-centric} I/O approach~\cite{pheromone} to optimize function communication via secure data objects, building on insights from prior work (e.g., CAP-VM~\cite{cap-vm}, Nephele~\cite{nephele}, Pheromone~\cite{pheromone}).
\projectname{}'s high function density increases co-location opportunities, and thus, it can also benefit from data-centric schedulers~\cite{abdi_palette_2023,jin_ditto_2023,pheromone} that effectively co-locate related functions.
% \projectname{} employs a \emph{data-centric} approach~\cite{pheromone} optimizing the function communication path.
% \projectname{} leverages \textit{function co-location} on the same host to create secure data objects in a CVM, building on insights from prior work (e.g., CAP-VM~\cite{cap-vm}, Nephele~\cite{nephele}, Pheromone~\cite{pheromone}) that utilize shared memory for efficient communication. 
% Co-locating functions in CVMs becomes challenging due to their limited scalability (\autoref{subsec:mot:server_consolidatiion}).
% \projectname{}'s high function density increases co-location opportunities, and thus, \projectname{} can also benefit from recent data-centric schedulers~\cite{abdi_palette_2023,jin_ditto_2023,pheromone} that effectively co-locate related functions. % , enabling its shared memory communication to be highly effective. 

In \projectname{}, \emph{data objects} serve as secure endpoints for trustlets to exchange data without traversing conventional I/O stacks.
The trusted monitor manages data object allocation and page table permissions, granting appropriate access to producers and consumers, and ensures secure data exchange while maintaining isolation between functions.
% When a function requires input, the trusted monitor allocates data objects and updates page table permissions -- granting write access to producers and read access to consumers -- ensuring secure data exchange while maintaining isolation between functions.
This design eliminates unnecessary cryptographic operations and redundant data copies, reducing I/O overheads. 
For functions that cannot be co-located, \projectname{} falls back to standard CVM I/O, ensuring compatibility with existing serverless frameworks.

\section{Design}
\label{sec:design}

\subsection{Trusted Monitor}

The trusted monitor is \projectname{}'s core component. It maintains strict isolation between each serverless function and the guest OS through hardware-enforced CVM partitioning and page table configurations. It operates at the highest privilege level of the CVM partitioning domains (e.g., VMPL-0). % and ensures the confidentiality and integrity of trusted processes.
% Further, delegating primary serverless management tasks to the guest OS reduces the software TCB.

\myparagraph{Trusted process management}
\projectname{}'s trusted monitor manages the CVM resources and preserves the state of the guest OS and trusted processes using process descriptors.
A process descriptor includes the values for general-purpose and control registers, address space information (page tables), process ID, CVM-partitioning privilege level, object information (\autoref{sec:design:chain}), and process state. % (i.e., created/initialized/ready/running/terminated).
The trusted monitor is responsible for allocating resources and scheduling the guest OS and the trusted processes. % by selecting the respective descriptor and switching the context to it.
Internally, the descriptor state is linked to one vCPU state of the CVM, and the trusted monitor uses a special instruction %(or TDX-module for TD-paritioning~\cite{td-partitioning})
to switch the context (e.g., AP creation call in SEV-SNP~\cite{sev-snp-abi}).
By executing the guest OS and the trusted processes at lower CVM partitioning levels, the trusted monitor governs their execution.

\myparagraph{Isolation enforcement}
%Unlike traditional nested virtualization~\cite{ben-yehuda_turtles_2010}, 
% The isolation among the trusted processes and the guest OS needs extra care due to the limited number of isolation domains (e.g., only four domains for VMPL~\cite{snp-white-paper}, three nested guests for TD-Partitioning~\cite{td-partitioning}).
The isolation between trusted processes and the guest OS requires careful design due to limited isolation domains (e.g., four domains for VMPL~\cite{snp-white-paper}, three nested guests for TD-Partitioning~\cite{td-partitioning}).
%This requires the trusted monitor to run multiple contexts in the same privilege domain. 
% This obliges the trusted monitor to run multiple contexts in the same privilege domain, sharing the same memory regions. % while ensuring isolation.
% To address this issue, the trusted monitor leverages the conventional ring protection, orthogonal to CVM partitioning: user-provided functions run at ring3, while the monitor-managed code runs at ring0, preventing manipulation of the trusted processes' page tables from ring3.
% The trusted monitor only inserts the minimal ring0 code, minimizing the possibilities of vulnerabilities in it, while it places the guest OS at a further lower privilege domain (i.e., VMPL-2). %, enabling strict isolation. 
To address this issue, the trusted monitor employs conventional ring protection, orthogonal to CVM partitioning: user functions run at ring3, monitor-managed code at ring0, preventing manipulation of the trusted processes' page tables from ring3, and the guest OS at a lower privilege domain (i.e., VMPL-2).
The page table-based isolation enables memory optimization through page sharing among functions—impossible in common CVMs—facilitating fast fork-based trustlet creation %from zygotes 
(\autoref{sec:design:zygote}) and efficient %trusted-monitor-mediated 
inter-function communication (\autoref{sec:design:chain}).

% This page table-based isolation allows the trusted monitor to optimize memory utilization by sharing commonly used pages among functions, which is impossible for the common CVMs.
% This choice realizes fast trustlet creation from a zygote by employing fork-based process creation (\autoref{sec:design:zygote}).
% Further, it allows the creation of efficient trusted-monitor-mediated communication paths between trusted processes (\autoref{sec:design:chain}).

\myparagraph{Programming interface}
As its execution environment base, \projectname{} adopts a LibOS architecture~\cite{baumann_bescule_2013,porter_libos_2011,cheriton_libox_1995,engler_exokernel_1995,lesile_mm_1996,gramine1}.
%The LibOS architecture aligns 
This aligns
well with the requirements of trusted processes, where the main code runs at the ring3, and offers users maximum flexibility through a common POSIX-compatible interface with a minimal resource footprint.
\projectname{} tailors the LibOS architecture for serverless execution by allowing the dynamic loading of functions without requiring the recompilation of other components. 
Furthermore, the LibOS provides a data-centric I/O~\cite{pheromone} API for inter-function communication (\autoref{sec:design:libos}) and relies on function provider policies that include workflow information (e.g., function chaining), measurements of the involved zygotes and functions, and, optionally, remote storage encryption keys for functions requiring external storage access.
%\projectname{} implements backend for the LibOS, including page table management and external I/O handling (\autoref{sec:design:libos}).

\begin{table*}[t]
   % \footnotesize
    \tabfontsize
    \centering
    \caption{Monitor system calls {\em define the interface between the guest OS and the trusted monitor (excerpt). }}
    % \vspace{-.2cm}
    \begin{tabularx}{\textwidth}{l|l|X}
         \hline
         {\bf Category} & {\bf API} & {\bf Description}  \\
         \hline
         \hline
         Zygote
         & createZygote(void* image) -> zHandle & Loads and attests an image of a zygote and returns a zygote handle. \\
         & deleteZygote(zHandle) & Removes the zygote and its derived trustlets. \\
         \hline
         Trustlet
         & createTrustlet(zHandle, fn) -> tHandle & Creates a trustlet based on a zygote and a provided function. \\
         & deleteTrustlet(tHandle) & Deletes the specified trustlet. \\
         & invokeTrustlet(tHandle, void*) -> Result & Run a function in a trustlet and return the result. \\
         \hline
         Attestation
         & attestMonitor() -> Report & Returns the attestation report of the monitor. \\
         & attest(Handle) -> Report & Retrieves  the attestation report of a trusted process. \\ 
         \hline
         Policy & loadPolicy(void*) & Loads the encrypted function provider policies. \\
         \hline
    \end{tabularx}    
    \label{tab:overview:api}
    % \vspace{-6mm}
\end{table*}

\myparagraph{Monitor system calls}
%The trusted monitor defines the monitor calls, which describe the interface between the trusted monitor and the guest OS or trusted processes (\autoref{tab:overview:api}).
%These monitor calls are divided into two main types: (1) for the guest OS to delegate requests for the execution of serverless functions, including creation of zygotes/trustlet, instantiating function execution, and creating trusted channels among trustlets and (2) for trusted processes for requesting backend operations, such as additional memory allocation and external file accesses (\autoref{sec:design:trustlet}).
A serverless framework, running within the guest, interacts with the trusted monitor through the defined monitor system calls (\autoref{tab:overview:api}).
These calls facilitate the execution of serverless functions, including the creation of zygotes/trustlets, function invocation, performing attestation, and loading the provider's policy.
%XXX: explain the detail of the policy%establishing trusted channels among trustlets.
Additionally, the trusted monitor defines an interface between trusted processes for the LibOS to request backend operations (\autoref{sec:design:libos}). %, such as additional memory allocation and external file accesses.
\revision{All interfaces are kept small and concise to eliminate the risk of possible attack vectors introduced by internal oversight.}

\myparagraph{Attestation service}
The trusted monitor includes an attestation service that generates an attestation report for remote users to verify the integrity of \projectname{} and its function execution.
%The guest OS obtains these reports via a monitor call.
On function execution, the attestation service calculates the measurement of function components %, including zygote, trustlet, function, and input/output,
%To realize faster report generation, the attestation service follows \projectname{}'s differential attestation principles (\autoref{sec:design:diffattest}), optimizing the measurement process by caching previously computed measurements.
following \projectname{}'s differential attestation principles (\autoref{sec:design:diffattest}) to optimize the measurement process by caching previously computed results.
%The root-of-trust of \projectname{} relies on CVM's attestation mechanism~\cite{sev-snp-abi, tdx-module}.
%Specifically, during \projectname{} initialization, the hypervisor loads the trusted monitor in the guest memory and sets an instruction pointer to its entry point.
%This initial guest state and memory are measured by the hardware and included in CVM's attestation report~\cite{sev-snp-abi}.
The root-of-trust of \projectname{} relies on CVM's attestation mechanism~\cite{sev-snp-abi, tdx-module}, whose attestation report includes the measurement of the initial guest state and memory.
The attestation service integrates this report into its final result. %, allowing users to verify the \projectname{}'s correct deployment.

\myparagraph{Performance consideration}
\revision{The trusted monitor is a central part of \projectname{}'s operations.
To ensure performant operations with multiple functions, the monitor allows each I/O operation to run simultaneously for multiple trustlets, without compromising its security properties.
%For memory accesses, the monitor is only called when memory needs to be allocated or deallocated, so that it can manipulate the trustlets' page table and set the correct VMPL levels for the new pages, an operation which can also be parallelized.
For memory accesses, the monitor is only called when (de)allocating memory. It manipulates the trustlets' page table and sets the correct VMPL levels for the new pages, an operation which can also be parallelized.
}

\subsection{LibOS Architecture}
\label{sec:design:libos}
\if 0
\projectname{} adopts a LibOS architecture for its trusted processes, intended to run in ring3.
%Precisely, a LibOS~\cite{baumann_bescule_2013,porter_libos_2011,cheriton_libox_1995,engler_exokernel_1995,lesile_mm_1996,gramine1} is implemented as a library that can be linked to applications and typically provides common OS services (e.g., filesystem). %., and allows for executing unmodified applications as libraries.
%In this setting, an application runs as a standard process, utilizing the LibOS's services instead of those of the host OS. 
A LibOS~\cite{baumann_bescule_2013,porter_libos_2011,cheriton_libox_1995,engler_exokernel_1995,lesile_mm_1996,gramine1} is a library offering common OS services to applications, enabling unmodified applications to run within a specific domain.
%
\fi
%\projectname{} adopts a LibOS architecture~\cite{baumann_bescule_2013,porter_libos_2011,cheriton_libox_1995,engler_exokernel_1995,lesile_mm_1996,gramine1} for its trusted processes, intended to run in ring3.
\projectname{}'s LibOS~\cite{baumann_bescule_2013,porter_libos_2011,cheriton_libox_1995,engler_exokernel_1995,lesile_mm_1996,gramine1} integrates a function runtime as well as a shim layer and filesystem with a nested namespace for efficient function execution.
%\projectname{}'s LibOS integrates a shim layer, which directly interacts with the trusted monitor via monitor system calls and a nested namespace for the file system to optimize the zygote image by placing certain files on the host while embedding commonly used files within the image itself.
%This approach reduces the image size, thus improving the boot time and enables fast file access to frequently used files.

%\masa{
%- explain service and components (cf. Gramine paper
%- memory layout (heap/stack) and allocation
%- how to handle I/O, interrupts, scheduling
%}

\myparagraph{Shim layer}
%In \projectname{}, 
The shim layer acts as the LibOS backend.
While the LibOS provides most required OS services, certain operations mandate assistance from the trusted monitor and guest OS.
The shim layer handles three key tasks: \emph{(i)}~memory allocation for trusted processes by coordinating with the trusted monitor, \emph{(ii)}~reading external host files, and \emph{(iii)}~handling data objects for function communication.
%Internally, it translates requests into monitor  calls.
%In both cases, LibOS delegates the operation to the shim layer, which translates requests from LibOS into monitor syscalls.
% The first is memory allocation beyond the initial memory allocated to the LibOS.
% In this case, the trusted monitor must allocate memory for the trusted process, extending the page table to cover that page. 
% The second is reading the external file on the host.
%For example, allocating additional memory that LibOS cannot perform independently.
% In these cases, LibOS delegates the operation to the shim layer, an interface that interacts with the trusted monitor.
% The shim layer, in turn, translates requests from LibOS into monitor calls.

% memo:
% - gramine PAL (shimlayer) manages gramine manifest, not the libos part
% - PAL checks the file hashes upon loading the file
% - for the hybrid fiels system, wallet embed files dirctly into the libos, not PAL

\myparagraph{Filesystem}
The LibOS provides an in-memory file system backed by a zygote, at the cost of a bloated image.
Removing non-commonly-used files reduces image size but compromises serverless function flexibility and user experience.
As a solution, \projectname{} employs a \emph{nested namespace}~\cite{joerg_cntr_2018} that integrates an embedded with an external filesystem.
From the application's perspective, all files appear at the same mount point (e.g., `/').
Internally, the LibOS first checks the embedded filesystem; if a file is missing, it retrieves it from the guest OS's filesystem via the monitor.
To ensure file integrity, the shim layer verifies files from the guest OS using provided measurements specified in a manifest~\cite{hunt_singularity_2007,tsai_gramine_2014} embedded in the zygote image and aborts operations in case of a mismatch.

% Q: who put the external files on to the guestOS?

% \pramod{still try to contrast it with conventional CVM networking.}
\myparagraph{Data-centric I/O}
%\pramod{add a network-specific LibOS details and also API based on peremone [NSDI 2023].}
% \dimitris{is Networking the right paragraph title here?}
For inter-function communication, the LibOS provides a \emph{data object} abstraction to the trusted processes.
% \autoref{tab:libos_api} shows the API for handling data objects. 
This abstraction allows a function to create, get, and update data objects via the API shown in \autoref{tab:libos_api}.
% Internally, \projectname{} employs a shared memory-based local object store~\cite{sabbioni_chain_2021,pheromone,kotni_faastlane_2021,qi_spright_2022,parola_sure_2024,chen_yuanrong_2024}, enabling a faster communication mechanism than the conventional network-based communication.
Unlike conventional CVM networking that requires traversing entire I/O stacks with multiple context switches and encryption overhead~\cite{cheng_tdxsurvey_2023, li_bifrost_2023,misono2024confidential, efficient_paravirt}, \projectname{} employs a shared memory-based local object store~\cite{sabbioni_chain_2021,pheromone,kotni_faastlane_2021,qi_spright_2022,parola_sure_2024,chen_yuanrong_2024}, enabling faster communication.
This is especially effective for \projectname{} as it avoids the costly CVM network overhead (\autoref{subsec:mot:network_communication}).
When the serverless orchestrator fails to co-locate functions, the trusted monitor follows a hybrid I/O approach~\cite{mahgoub_sonic_2021,pheromone} and resorts to the standard CVM I/O primitives via the guest OS, similar to the external file system access described above.
Importantly, this mechanism ensures transparent communication between functions, regardless of whether they are co-located or distributed.
\autoref{sec:design:chain} details \projectname{}'s communication mechanism.

\subsection{Trusted Process}
\label{sec:design:trustedprocess}

%\pramod{tie it with the libOS, i.e., a trusted process encapsulates the libOS. i think this figure can be referred in prev section s5.2}

\begin{figure}[t]
     \centering
    \includegraphics[width=\columnwidth]{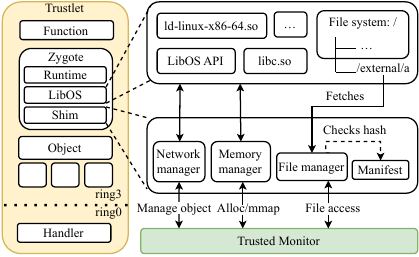}
     % \vspace{-2mm}
     \caption{Trusted process architecture.
     %\textit{A trustlet is created from a template zygote image, which includes LibOS and function runtime. Each trustlet loads its own function and has an input and output channel for efficient data transfer.}
     }
     \label{fig:trustlet}
     %\Description[<short description>]{<long description>}
     % \vspace{-3mm}
\end{figure}

%\autoref{fig:trustlet} shows the architecture of a trusted process.
A trusted process (\autoref{fig:trustlet}) is designed to efficiently host serverless functions within an isolated domain. % in a lightweight manner.
Each trusted process encapsulates the LibOS, which is independent of the guest OS and provides functionalities to host a serverless runtime.
Trusted processes run within the same CVM privilege domain, enabling memory sharing and CoW-based process creation.
Meanwhile, the trusted monitor ensures their strict isolation by controlling the per-process page tables.
Trusted processes are categorized into two types: \emph{zygotes} and \emph{trustlets}.

%In \sysname, each serverless function is executed as a \emph{trusted process} (\autoref{fig:trustlet}).
%The trusted monitor prevents the guest OS from accessing the memory of trusted processes using VMPL configurations. 
%To ensure isolation between trusted processes, \sysname leverages ring protection and per-process page tables~\cite{veil_ahmad_2024}, entirely managed by the trusted monitor.

\begin{table}[t]
\centering
\tabfontsize
\caption{LibOS API for data objects.}
% \vspace{-.2cm}
\begin{tabular}{l|l}
 \hline
  \textbf{API} &  \textbf{Description} \\
  \hline
  \hline
  {createObject}(\textit{len,type}) $\rightarrow$ \textit{obj\_id} & Create data object.\\ %(type=shared/pipe). \\
  %& (type=shared or pipe (exclusive write)) \\
  \hline
  {getObject}(\textit{obj\_id}) $\rightarrow$ \textit{*obj} & Get a pointer to the object \textit{obj\_id}. \\
  %getValue(\textit{obj\_id}) $\rightarrow$ \textit{value} & Get a value of object \textit{obj\_id}.\\
  %setValue(\textit{obj\_id, *value}) & Set a value to object \textit{obj\_id} \\
  \hline
  {getInputObject}() $\rightarrow$ (\textit{obj\_id, len}) & Get data object that contains input.\\
  {setOutputObject}(\textit{obj\_id}) & Set data object as output. \\
  \hline 
\end{tabular}
\label{tab:libos_api}
% \vspace{-3mm}
\end{table}

\subsubsection{Zygote}
\label{sec:design:zygote}
A \emph{zygote} is a template process containing a function execution environment tailored for a serverless runtime.
% As a term, it was first coined by Android~\cite{zygote,lee_morula_2014}.
%In \projectname{}, zygotes adopt a LibOS architecture (\autoref{sec:design:libos}), hosting the serverless runtime and essential system services, running in ring 3 within a single, isolated address space.
%This is due to the fact that trusted processes operate in a different memory space (VMPL level) isolated from the guest OS and cannot rely on its functionalities. 
%Importantly, zygotes do not execute functions but rather serve as a base for the instantiation of separate function execution contexts, namely trustlets (\autoref{sec:design:trustlet}).
Rather than running functions itself, it serves as a base for the instantiation of separate function execution contexts, namely trustlets (\autoref{sec:design:trustlet}). 
They are fetched from the zygote registry, and after the successful integrity verification by the attestation service, 
%the process manager allocates memory and loads the image, initializing their process descriptor and page tables.
the trusted monitor initializes their process descriptor and page tables.
%Further, the monitor sets the zygote memory to read-only and sets up its page tables that restrict access only to its memory, ensuring isolation among trusted processes, similar to prior work~\cite{veil_ahmad_2024}. 
%This instantiation employs a copy-on-write-based forking mechanism~\cite{sand,catalyzer,sock,du_xpusim_2022,li_pagurus_2022} that allows for fast function startups and efficient memory utilization.
To optimize the function invocation process, a zygote offers a pre-initialization mechanism, enabling it to preload and initialize the function runtime environment~\cite{seuss}.
During the zygote loading phase, the trusted monitor executes the zygote, and the zygote notifies about the completion of the initialization by making a specific monitor call.
The trusted monitor then seals the initialized state and marks all zygote memory non-writable for the subsequent creation of trustlets.

%\myparagraph{Pre-initialization}

% \myparagraph{zygote creation}
% \dimitris{below paragraph is a repetition / merge the important details above}
% To create a zygote, the guest OS initially fetches an image from the registry and invokes the createZygote() monitor call. 
% Then, the attestation module attests the zygote image using the function provider's policy that contains its measurement to verify its integrity (\autoref{sec:design:diffattest}).
% Subsequently, the process manager allocates memory for the zygote and loads its image.
% Further, the monitor sets the zygote memory to read-only and sets up its page tables that restrict access only to its memory, ensuring isolation among trusted processes, similar to prior work~\cite{veil_ahmad_2024}. % they share the same physical pages at the same VMPL level.
% The page table only allows it access to the corresponding zygote image memory.
% This page table enforces isolation among trusted processes, 

\subsubsection{Trustlet}
\label{sec:design:trustlet}
A trustlet serves as the execution environment for serverless functions, inheriting the memory layout and execution environment from its base zygote. Each trustlet has data objects for inter-function communication. On trustlet creation, the trusted monitor duplicates the zygote's process descriptor, loads the function, and inputs data to the corresponding memory region—all without copying the zygote memory.%, realizing fast function startup.

\myparagraph{Copy-on-write mechanism}
Any trustlet \emph{write} attempt to the non-writable zygote memory results in a page fault.
To handle page faults, the trusted monitor installs a page fault handler for each trustlet.
This handler delegates the actual page fault handling to the trusted monitor via a monitor call.
Subsequently, the trusted monitor copies the page, updates the trustlet's page table, and then resumes the execution.
%If the reserved VMPL-1 memory is not sufficient, it allocates memory from VMPL-0, and converts it to VMPL-1 using RMPADJUST.
%Next, the runtime copies the data to the newly allocated region and updates the trustlet's page table to include the page with the write permissions. % to point to it with writable permission.
%Finally, the runtime resumes the execution of the trustlet.

\myparagraph{Input and output}
Function input and output are represented as special types of data objects.
A function interacts with these objects via the API in \autoref{tab:libos_api}.
When invoking a trustlet, the trusted monitor creates an object for its input and retrieves the results from its output object after its execution.

\myparagraph{Scheduling}
The guest OS initiates the scheduling of trustlets via the invokeTrustlet() call. %, which hands the control to the process runtime to manage the trustlet execution.
% Namely, invoketrustlet() monitor call schedules the trustlet.
% Internally, the process runtime manages the scheduling of trustlets.
\projectname{} runtime enforces a run-to-completion execution model, i.e., \projectname{} executes a trustlet until completion for efficient execution~\cite{serverless_scheduler_costs}, and returns the result to the guest OS. 
\revision{This fits for serverless workloads, which are most of the time short-lived. In case a function performs I/O tasks, the trustlet returns control back to the monitor and a different trustlet can be scheduled, even before the original one has finished its execution.}
On top of that, the scheduler is also responsible for handling the function chaining of the co-located trustlets (\autoref{sec:design:chain}).

\subsection{Confidential Network Architecture}
\label{sec:design:chain}
\begin{figure}[t]
     \centering
     \includegraphics[width=\columnwidth]{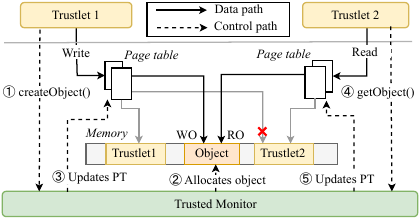}
     % \vspace{-1mm}
     \caption{Inter-function communication with data objects.}
     \label{fig:libos:networking}
     %\Description[<short description>]{<long description>}
     % \vspace{-3mm}
\end{figure}

%\pramod{MAKE IT DETAILED here. Explain function chaining as a special case for the network architecture.}
\myparagraph{Trusted-monitor-mediated I/O path}
Data objects constitute a fundamental component of the inter-function communication in \projectname{}.
\autoref{fig:libos:networking} illustrates their control and data path.
In this example, Trustlet1 sends data to Trustlet2 via a data object.
First, Trustlet1 invokes the {createObject}() API call (\Circled{1}).
Internally, the trusted monitor allocates the object (\Circled{2}) and updates the Trustlet1's page table to grant it write access (\Circled{3}).
% With this permission, Trustlet1 can write to the data object.
Subsequently, when Trustlet2 gets the object via {getObject}() (\Circled{4}), the trusted monitor updates the Trustlet2's page table, granting it read permission (\Circled{5}).
Through this trusted monitor-mediated shared-memory communication mechanism, \projectname{} establishes a secure, yet efficient, channel between functions while maintaining strict isolation. 
\revision{Further, the monitor ensures that each object is attached to only two trustlets at most, one with read and one with write access.}

\myparagraph{Function chaining}
Chaining multiple functions is a common pattern in serverless computing~\cite{daw_xanadu_2020,tariq_sequoia_2020,serverlessbench}.
% Thus, supporting efficient communication mechanisms is crucial. 
\projectname{}'s data object enables efficient function chaining through shared memory~\cite{sabbioni_chain_2021,qi_spright_2022}.
Specifically, when chained functions are scheduled on the same host, the trusted monitor creates a special data object and grants write permission to the first function and read permission to the second function.
If the functions can not be co-located, \projectname{} falls back to the normal network path: it returns the result of the first function to the serverless orchestrator, and then the orchestrator invokes the second function.

\revision{Note that, these chains are not limited to two functions and could potentially result in a circular chain, leading to the functions not producing a result.
The current \projectname{} prototype does not consider this case. However, circular chains can be detected and prevented by the monitor in a future
version. or \projectname{}’s policies could be extended to explicitly disallow it.}

\revisionmyparagraph{External network I/O}
\revision{
To facilitate network communications in \projectname{} across the CVM boundaries, additional steps are required.
%The communication is not directly done by \projectname{}'s monitor; 
%instead, the communication needs to be handled externally.
The monitor delegates the external network I/O handling to the guest OS.
%A serverless framework could handle the result, which has been encrypted by the current monitor instance, and forward it to the desired \projectname{} instance. 
%If inter-server communication is required, the serverless framework must handle it appropriately.
In practice, the monitor should encrypt the network data and provide it to the serverless framework for forwarding.
Then, the serverless framework can send this data to the desired destination and get a response. %that can decrypt it and proceed with further processing.
}

\begin{figure}[t]
     \centering
     \includegraphics[width=\columnwidth]{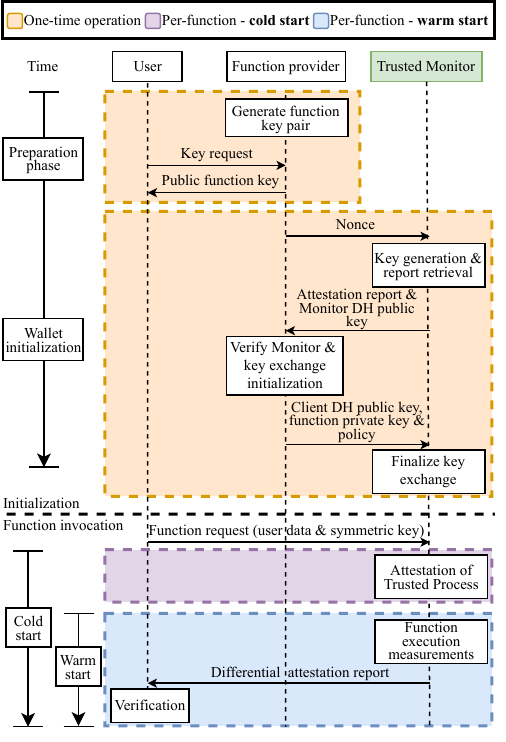}
     % \vspace{-2mm}
     \caption{Workflow of \projectname{}'s differential attestation.
    }
     \label{fig:attestation}
     % \vspace{-3mm}
\end{figure}

\subsection{Differential Attestation}
\label{sec:design:diffattest}

In confidential serverless computing, besides attesting the underlying platform, users need to verify the serverless runtime, the deployed function, and any inputs and outputs to validate the correct execution. 
This process can severely impact the function's startup and execution latency.
To alleviate this issue, \sysname introduces a \emph{differential attestation protocol} to reduce the latency of verifying serverless function execution, which requires attesting the platform, runtime, function, and I/O. 
The key idea is to retain previously calculated measurements 
and compose a cumulative attestation report by measuring only the mutable parts at runtime for the function execution. 

\myparagraph{Security guarantees}
\projectname{}'s differential attestation ensures the integrity of the monitor, trustlet, zygote, function, its input, and output.
It relies on the correctness of \projectname{}, in particular on \projectname{}'s monitor, which is the only permanently immutable component of the system.
All other parts of the differential attestation are at one point mutable, since they are fetched and instantiated by the monitor.
After their initialization, all parts that constitute a trustlet become immutable with data protected with CoW. 
%After their initialization, all parts that constitute a trustlet become immutable, with the zygote only being accessible via CoW and the function code being read-only.
This enables the monitor to cache the hashes of these components for future use. 
% The hashes are based on the initial file content (e.g., the ELF file of the Zygote).

\myparagraph{Attestation flow}
\autoref{fig:attestation} illustrates the workflow of \projectname{}'s differential attestation protocol. It consists of two main phases: the \emph{initialization} and the \emph{function invocation} phase. 

The \textbf{initialization} phase begins with the function provider sharing the public function key with the user, which is used to encrypt the user requests.
Then, during the \sysname initialization phase, the function provider establishes a secure connection with the \sysname instance to share the function's private key and the function's policy that specifies the measurements of zygotes and functions permitted for use.
The establishment of the secure connection relies on the attestation of the trusted monitor, whose trust is rooted in the ASP~\cite{sev-snp-abi}.

Precisely, the function provider first sends a nonce to the trusted monitor to prevent replay attacks. 
In turn, the attestation service generates a Diffie–Hellman (DH) key pair and retrieves the attestation report from the ASP, 
which includes the measurement of the trusted monitor, the provided nonce and the hash value of the DH public key as user data. 
The function provider receives and verifies the report and then sends the function private key and the policy to the \sysname instance.
\sysname uses the function’s private key to decrypt user requests and the policy to attest the trusted processes.

Importantly, before providing private or security-sensitive data as a function input, a user must first perform a test function call, retrieve the attestation report, and verify that the \projectname{} instance is at the expected state and the correct toolchain is present. After this point, the user has gained trust in \projectname{} and can proceed with their function invocations.

The \textbf{function invocation} phase includes two scenarios depending on the \sysname's state, namely \emph{cold} and \emph{warm} start (\autoref{fig:invocation_phases}).
In the cold start case, the attestation service verifies the integrity of the trusted process by comparing the measurements of its components against the values provided by the function provider's policy during the initialization phase.
Beyond this point, 
subsequent invocations of the same function become a warm start. \sysname's attestation service only has to measure the function input and output and can generate a report based on prior measurements of the monitor, the zygote, and the invoked function.
When several serverless functions are chained, the extended attestation report includes all their measurements.
This approach reduces the attestation time while ensuring verifiability for the entire data chain.

\section{Security Analysis}
% \dimitris{make a summary basically and mention that analytical description is provided in the supplementary material}

\if 0
% \subsection{Security Analysis of the trusted monitor}
In this section, we present potential attack vectors, how \sysname mitigates them (\autoref{tab:security_analysis}), and the formal verification of \projectname{}'s security protocols. 
Due to space constraints, more details on \projectname{}'s security analysis, as well as the full proofs and modeling details, are available in \autoref{appendix:sec_analysis}.
\fi

\begin{table}[t]
\centering
% \footnotesize
\tabfontsize
\caption{Potential attack vectors and \sysname's mitigations.}
%\textit{\projectname offers confidentiality and integrity protection against attacks from the outside of the CVM (from the host), inside the CVM (guest OS), and from the co-located functions.}}
% \vspace{-.2cm}
\begin{tabular}{l|l} 
\hline
\textbf{Attack vector} & \textbf{Mitigation} \\
\hline
\hline
\textbf{From guest OS} &  \\
~~Read the \sysname's memory & VMPL protection \\
~~Read user data in the request  & Data encryption \\
~~Load crafted zygotes/functions/inputs & Attestation \\
~~Load compromised external file & Measurement check \\ % This hash checking is done in the shim layer, not the trusted monitor
\hline
\textbf{From host and hypervisor} &  \\
~~Access VM's memory & CVM protection \\
~~DMA to the VM's memory & CVM protection \\
~~Inject malicious interrupts~\cite{schlüter2024wesee,schlüter2024heckler} & Alternate injection~\cite{sev-snp} \\
~~Return invalid CPUID values~\cite{li_teesok_2024} & CPUID pages~\cite{sev-snp-abi} \\
~~Launch compromised images & Attestation \\
~~Manipulate user request & Protocol encryption \\
\hline
\textbf{From co-located functions} &  \\
~~Access other trustlet's memory & Page table protection \\
~~Reuse-based attacks~\cite{zhao_resusableenc_2023} & Recreate trustlets \\
\hline
\end{tabular}
\label{tab:security_analysis}
% \vspace{-2mm}
\end{table}

\subsection{Attack vectors \& \sysname's mitigations}
\autoref{tab:security_analysis} summarizes potential attack vectors and \projectname{}'s mitigations.

\myparagraph{Guest OS}
First, we consider attacks originating from the untrusted guest OS within the CVM.
An attacker may attempt to access data of a serverless request. However, this data is encrypted with a user key that only the trusted monitor possesses and is inaccessible to external entities, provided that the user secrets are not compromised.
On top of that, modifying the VMPL configuration via the RMPADJUST instruction~\cite{sev-snp-abi} is prohibited in the guest OS (VMPL-2).
Thus, any attempt from the guest OS to access memory regions of the trusted monitor and Trusted Processes fails thanks to VMPL protections, which are set by the trusted monitor and trap such accesses.

Further, an attacker could load arbitrary zygotes and functions.
However, \sysname's attestation service blocks this process if there is a measurement discrepancy from the values in the function provider's policy. 
Replay attacks by replacing user input are also detected as users can validate the integrity of the input based on the content of the attestation report.
Lastly, if an attacker intercepts a file request from a trustlet to the nested filesystem (\autoref{sec:design:libos}) and returns a compromised file, the LibOS verifies the file's measurement against its expected value in \sysname's manifest, rejecting any tampered files. 

\myparagraph{Host and hypervisor}
CVMs employ mechanisms to protect CVM's private memory and provide ways to retrieve trusted system information from within the CVM.
Precisely, unauthorized access to CVM memory is blocked by AMD SEV-SNP’s reverse map page table (RMP), which prevents the host, hypervisor, and other co-located VMs from accessing CVM's private memory. 
Direct memory access (DMA) attacks from the host are also infeasible since DMA is prohibited by hardware. 
Further, to address potential threats to compromise the guest via arbitrary interrupt or exception injection~\cite{schlüter2024wesee,schlüter2024heckler}, SEV-SNP introduces alternate interrupt injection~\cite{sev-snp}, restricting the hypervisor to specific exception types. 
On top of that, an attacker may try to disrupt the system by tampering with the CPUID instruction~\cite{li_teesok_2024} return values.
As a countermeasure, SEV-SNP provides a CPUID-page mechanism, which ensures correct CPUID values, validated by the ASP.

Attempts to compromise \sysname by launching a tampered trusted monitor, bypassing it altogether, or directly booting \sysname in a standard VM are detected during \sysname's remote attestation workflow. Attestation reports include invalid measurement values if altered software is used. Attempts to deploy \sysname as a normal VM can be identified, as, in this case, \sysname fails to produce a valid report from the ASP. Even crafted reports are detected, since only the CPU vendor’s key, securely managed by the ASP, can sign genuine attestation certificates. Lastly, establishing secure connections (e.g., TLS) protects against network communication manipulation by ensuring the confidentiality and integrity of network packets.

\myparagraph{Co-located serverless functions}
\sysname prevents attacks from co-located functions.
An attacker might create a serverless function that exploits a runtime vulnerability aiming to access memory belonging to another trustlet.
However, each trustlet runs in ring3 and has its own page tables managed by the trusted monitor, which prohibits access to other trustlet's memory.
Further, to prevent reuse-based attacks~\cite{zhao_resusableenc_2023}, \sysname's process runtime recreates trustlets when receiving a request from a different user, resetting the potentially compromised residual state due to a previous invocation.

\subsection{Formal verification of Protocols}
We formally verify \projectname{}'s differential attestation protocol, as depicted in \autoref{fig:attestation}, using the Tamarin Prover~\cite{tamarin_paper, tamarin_site} under the Dolev-Yao~\cite{dolev_yao} attacker model. 
The verification considers two key properties for \sysname's differential attestation protocol: \emph{(i)}~\textbf{secrecy}, secrets (e.g., encrypted data, private keys) remain undisclosed unless explicitly compromised; and \emph{(ii)}~\textbf{authenticity}, guaranteeing that attested results correspond to correct function execution with validated inputs. 
We model the steps specific to \sysname in detail.

More precisely, an unbounded number of protocol flows are considered simultaneously, with their individual states represented as part of a global multiset. State transitions are encoded as rewriting rules, which are one of the input formats for the Tamarin prover, and long-term secrets are treated as compromisable.
Messages are considered atomic, and cryptographic functions (e.g., hashing, encryption) are assumed to be perfect, i.e., without side effects or collisions, and known to the attacker.
We further assume the correct functionality of the underlying hardware, i.e., the ASP does not leak its secrets and produces correct reports, and report verification infrastructure. Such external dependencies are treated as a black box.

Our model is based on \emph{action facts}, which denote events in the protocol trace.
We use the \textit{Secret(s)} action fact to explicitly mark secret information \textit{s}, in conjunction with the builtin \textit{K(x)} action fact, which marks an adversary obtaining information \textit{x}.
We further introduce the \textit{UsrTrustRes(...)} action fact to indicate that a user trusts, after the last protocol step, that the obtained result was indeed computed on a trusted software stack.

\myparagraph{Rules}
To translate the protocol specification to rules, which operate on a global state, we identify: 
\emph{(i)} the necessary inputs from the network, as well as, the persistent states for each Agent for the rules left-hand side, \emph{(ii)} the resulting outputs on the network, as well as, any modifications to the persistent states of each Agent for the rules right-hand side, \emph{(iii)} any checks performed by an Agent, which may translate to restrictions of the rule transitions.

This approach enables a systematic derivation of the model from the specification for most of the protocol, some parts, however, require additional consideration:

\textbf{The preparation phase} contains an exchange of the public function key, between the user and the function provider, which is not further specified, for the formal analysis we model this exchange as a secure channel (\autoref{appx:secure-channel-rules}). The motivation for this is, that we do not consider attacks on this information exchange. The attacker is still able to obtain the function's public key, which we model with \\ \textit{Out(pk(\raisebox{0.5ex}{\texttildelow}func\_priv))} on the right-hand side of the key generation rule, but unable to provide the user false information in the preparation phase.

\begin{figure}[h]
\vspace{0.5em}
\footnotesize
\begin{verbatim}
rule ChanOut_S:
    [ Out_S($A,$B,x) ] 
    --[ ChanOut_S($A,$B,x) ]->
    [ !Sec($A,$B,x) ]
    
rule ChanIn_S:
    [ !Sec($A,$B,x) ] 
    --[ ChanIn_S($A,$B,x) ]->
    [ In_S($A,$B,x) ]
\end{verbatim}
\vspace{-1em}
\caption{Secure channel rules from the Tamarin Manual\cite{tamarin_manual}.}\label{appx:secure-channel-rules}
\end{figure}

\textbf{The initialization phase} is largely based on Tamarin's \textit{diffie-hellman} builtin, for modelling the key exchange. To model the report generation, verification and user-data retrieval as a black-box we introduce the functions \textit{gen}, \textit{verif} and \textit{getD} respectively. The expected behaviour of these is expressed in the following equations, where \textit{m} refers to the machine, \textit{d} to the software measurement, and \textit{u} to the retrievable user-data:

\begingroup
\setlength{\abovedisplayskip}{-0.7em} 
\setlength{\belowdisplayskip}{0.1em} 
\setlength{\jot}{0.1em} 
\begin{equation}
    \begin{split}
           verif(gen(m, d, u), m, d) = true
    \end{split}
\end{equation}
\begin{equation}
    \begin{split}
           getD(gen(m, d, u)) = u
    \end{split}
\end{equation}
\endgroup

In this phase we also included an unrestricted \textit{machine\_init} to be able to set up an infinite number of machines and thus infinite potential \projectname{} instances. This, together with the way we modeled the function provider in the previous phase, allows us to model an unbounded number of \projectname{} instances, machines, functions, providers, and users all while ensuring they are unable to interfere to violate our desired properties.

The remainder of the rules are straightforward translations from the protocol.

\myparagraph{Lemmas}
We use \(\textit{a} ~@~ t_i\) to denote that action fact \(\textit{a}\) occurred at time \(t_i\).
Leveraging these action facts, we verify the following lemmas for \sysname's differential attestation protocol:
\begin{itemize}[wide, labelindent=0pt, nosep]
\item \textbf{Secrecy lemma:} If some data (e.g., function input/output) is declared as secret \textit{s}, it remains undisclosed to attackers unless one agent \textit{A} having access to it (\textit{KnowSecret(A, s)}) is explicitly compromised (\textit{Compr(A)}).
\begingroup
\setlength{\abovedisplayskip}{0.1em} 
\setlength{\belowdisplayskip}{0.1em} 
\setlength{\jot}{0.1em} 
\begin{equation}
    \begin{split}
           \forall~\textit{s}, t_i~.~\textit{Secret(s)} @t_i &\implies \nexists~t_j~.~\textit{K(s)} @t_j ~\vee~ \\           (\exists~\textit{A},t_r~.~\textit{Compr(A)}@t_r & ~\wedge~ \textit{KnowSecret(A, s)}@t_i)
    \end{split}
\end{equation}
\endgroup
\item \textbf{Authenticity lemma:} 
If a user trusts a result, after verifying the differential attestation report, then the result \textit{res} was computed on the trusted monitor in a valid state \textit{tm}, and with the expected zygote \textit{zy}, trustlet \textit{tr}, and input \textit{in}, which is modeled by the \(compute\) function.
\begingroup
\setlength{\abovedisplayskip}{0.1em} 
\setlength{\belowdisplayskip}{0.1em} 
\setlength{\jot}{0.1em} 
\begin{equation}
    \begin{split}
           \forall~&\textit{tm}, \textit{zy}, \textit{tr}, \textit{in}, \textit{res}, t_i~.~\textit{UsrTrustRes(tm,zy,tr,in,res)} @t_i \\
           &\implies compute(\textit{tm},\textit{zy},\textit{tr},\textit{in}) = \textit{res}
    \end{split}
\end{equation}
\endgroup
\end{itemize}

Tamarin verifies the specified properties by showing that there is no trace that leads to a falsification of these lemmas.
Thus, we show that the attestation protocol ensures secrecy, authenticity, and correctness under the specified assumptions.

However, to increase confidence in the correctness of our model we include a simple sanity check, which ensures that it is possible to reach this point in the protocol. For this we use this lemma:

\begingroup
\setlength{\abovedisplayskip}{-0.7em} 
\setlength{\belowdisplayskip}{0.4em} 
\setlength{\jot}{0.1em} 
\begin{equation}
    \begin{split}
           \exists~&\textit{tm}, \textit{zy}, \textit{tr}, \textit{in}, \textit{res}, t_i~.~\textit{UsrTrustRes(tm,zy,tr,in,res)} @t_i
    \end{split}
\end{equation}
\endgroup

If this lemma is falsified, then the \textit{authenticity lemma} would trivially hold, since it would be a for-all condition on an empty set. However, this would likely indicate an issue in our model and not the proper functioning of the protocol, therefore we include the above sanity check to rule out this edge case. 

\begin{table}[t]
\centering
\footnotesize
\caption{Agents with access to secret information}
\begin{tabular}{l|l} 
\hline
\textbf{Information} & \textbf{Agents with access} \\
\hline
\hline
User's Function input & User, TrustedMonitor, Function Provider \\
Output encryption key & User, TrustedMonitor, Function Provider \\
Function's private key & TrustedMonitor, Function Provider \\
\hline
\end{tabular}
\label{appx:secrets}
\end{table}

For the secrecy analysis, we were also required to explicitly state the agents that can potentially access, i.e., obtain in plain text, certain secret information, as depicted in \autoref{appx:secrets}. We found the sets of agents to be minimal, as further reduction caused the \textit{secrecy lemma} to be violated. Notably, this required us to include the \textit{Function Provider} as an agent, with potential access to the User's function input and output. The reason for this is the private function key, which is known to the function provider. 

The persistence of these long-term keys and their use also prevented us from verifying a stronger secrecy property called \textbf{perfect forward secrecy}. This would ensure that information will remain secret, even if the attacker is able to record the communication and \textit{compromize} some long-term key in the future. Checking for this property entails making a small modification to the \textit{secrecy lemma}:

\begingroup
\setlength{\abovedisplayskip}{-0.5em} 
\setlength{\belowdisplayskip}{0.2em} 
\setlength{\jot}{0.1em} 
\begin{equation}
    \begin{split}
           \forall~\textit{s}, t_i~.~\textit{Secret(s)} @t_i &\implies \nexists~t_j~.~\textit{K(s)} @t_j ~\vee~ \\           (\exists~\textit{A},t_r~.~\textit{Compr(A)}@t_r & ~\wedge~ \textit{KnowSecret(A, s)}@t_i \bm{~\wedge~ t_r < t_i})
    \end{split}
\end{equation}
\endgroup

For our protocol, Tamarin is able to correctly identify that this lemma is violated since compromising the long-term private key of a function enables the decryption of any previous request to that function. We consider this attack vector out of scope for this paper, as it relies on \textit{compromising} a key component of our infrastructure. Future work may explore extensions to the protocol, that address this, to better contain an initial compromise.
\section{Implementation}
\label{sec:implementation}

\myparagraph{Trusted monitor}
We implement our prototype of \sysname for Linux environments.
We base \sysname's trusted monitor on COCONUT-SVSM~\cite{coconut_svsm}.
COCONUT-SVSM is a service module that operates at VMPL-0 and aims to provide services (e.g., vTPM~\cite{narayan_vtpm_2023}) to a guest running at VMPL-2.
We extend COCONUT-SVSM to support trusted processes in VMPL-1 and incorporate \sysname's differential attestation service.
We base our LibOS for trusted processes on Gramine~\cite{gramine1,gramine2} and implement a backend to enable its execution at VMPL-1.
\sysname's current prototype uses a patched Linux version v6.8 (host OS) and v6.5 (guest OS), configured to support COCONUT-SVSM.

\myparagraph{Communication protocol}
The COCONUT-SVSM implements a guest communication protocol defined by AMD~\cite{sev_svsm_spec}, enabling the guest OS to request services from the COCONUT-SVSM using the VMGEXIT instruction.
We extend this protocol to support the monitor system calls (\autoref{tab:overview:api}) by adding a new service type and implementing its corresponding handlers.

For communication between the trusted monitor and trustlets, we enable \#VC (VMM Communication Exception) reflection~\cite{sev-snp-abi}, allowing the trusted monitor to trap \#VC during trustlet execution. 
We define a communication protocol based on the cpuid instruction, leveraging the unused hypervisor CPUID Leaf range~\cite{amd-sdm}.
Precisely, trustlets set a service type in the RAX register and execute the cpuid instruction, triggering \#VC.
In turn, the trusted monitor inspects the RAX value, executes the corresponding monitor service, if applicable, or the normal cpuid instruction, and returns the result.

\myparagraph{LibOS}
Our LibOS for trusted processes uses Gramine~\cite{gramine1,gramine2} with a backend for VMPL-1 execution.
Gramine defines a Platform Adaptation Layer (PAL) host ABI~\cite{gramine-pal} to delegate specific operations to the host. 
We implement our PAL as a shim layer for our trusted monitor, which allows \sysname to run the Python runtime as a trusted process.
Its core functionalities include memory allocation and external file access, and it leverages \sysname's communication protocol to perform its PAL ABI operations requests.

For memory allocations, the trusted monitor handles the requests from the PAL. Specifically, it allocates memory, adjusts the VMPL access permissions using the RMPADJUST instruction, and updates the trustlet's page table entries.
For external file access, the trusted monitor returns a special value to the guest OS to request the desired file.
Then, the guest OS reads the file and invokes the invoketrustlet() monitor call.
Subsequently, the trusted monitor copies the file data into the trustlet's memory and resumes its execution.
The trustlet validates the file integrity before using it by comparing its measurement with a specified value in the manifest.
For the memory backend, the trusted monitor secures the memory pool and validates each page during its boot to optimize memory allocation at runtime.
When a trustlet requests memory, the trusted monitor allocates memory from the pool, and updates the trustlet’s page table entries accordingly.

\myparagraph{Copy-on-write handling}
Gramine LibOS operates in ring3 and does not contain exception handling code.
To realize copy-on-write-based forking, we develop a minimal exception handler that executes in ring0 within a trustlet (VMPL-1).
The trusted monitor configures the GDT (Global Descriptor Table), IDT (Interrupt Descriptor Table), and TSS (Task State Segment) of the trustlet to ensure that it uses the handler upon launch.
On a page fault, the page fault handler delegates the actual handling to the trusted monitor via a monitor call.
The trusted monitor resolves it by updating the page table entries appropriately and then resumes the execution of the trustlet.

\myparagraph{Attestation service}
\sysname leverages COCONUT-SVSM's \texttt{get\_regular\_report()} to retrieve a signed AMD SEV-SNP report and, then, stores it securely in the trusted monitor's memory. To calculate the measurements of individual trusted processes components (e.g., zygote image, trustlet), \sysname uses the SHA-512 algorithm (same as the SEV-SNP's measurement algorithm~\cite{sev-snp-abi}). These measurements are cached in the zygote and trustlet contexts for faster retrieval. The monitor invokes the attestation functions internally when fetching a zygote or creating a trustlet, validating them against the function provider's policy by comparing the calculated with the provided measurements.  
Finally, on a function invocation, the measurements of the used software components and the function input and output are appended to the trusted monitor's attestation report, which is then signed using the private key provided by the function provider and forwarded to the user for verification of the entire execution.

\myparagraph{Guest OS components}
\projectname{} includes a kernel module that allows a guest application to make monitor calls.
In addition, we implement a Python library that interacts with the trusted monitor through this kernel module.
We also extend the SeBS benchmark suite~\cite{copik2021sebs} to support running serverless functions with \projectname{} using this Python library.

\section{Evaluation}
\label{sec:evaluation}

We evaluate \projectname{} by analyzing its end-to-end performance (\autoref{subsec:performance}),the performance of its operations (\autoref{sec:eval:analysis}), its resource efficiency (\autoref{sec:eval:analysis:memory}), its communication network performance (\autoref{subsec:networking-eval}), and its scale-out capabilities (\autoref{subsec:traces}).% under production-like workloads .
%Alternate interrupt injection is also missing but planned for future integration as COCONUT-SVSM develops it. Full integration with serverless orchestration frameworks is future work.

\subsection{Experimental Setup}

\myparagraph{Testbed}
We perform our experiments on an AMD SEV-SNP-enabled server with an AMD EPYC 7713P CPU (64 cores, hyperthreading disabled) and 1024 GB of DDR4 DRAM (16$\times$64 GB/DIMM).
The server runs NixOS 24.11 with an AMD SEV-SNP-enabled Linux kernel (v6.8.0).
Each VM uses an Ubuntu-22.04 with a VMPL-enabled Linux kernel (v6.5.0).

\myparagraph{Variants}
We use the baselines summarized below.% \autoref{tab:variants}. 
\begin{table}[H]
\centering
% \footnotesize
\tabfontsize
%\caption{Benchmarking variants.}
\begin{tabular}{c | l } 
 \hline
 \textbf{Variant} & \textbf{Execution environment} \\
 \hline
 \hline
 Native & Bare-metal instance on Linux \\
 \hline
  LibOS (Gramine) & Gramine LibOS on Linux\\
 \hline
 Container (Kata) &  Kata containers runtime with QEMU/KVM\\
 \hline
 VM (KVM-Linux) & Standard VM with Linux guest OS \\
 \hline
% Firecracker & microVM based on Amazon FireCracker \\ 
% \hline
CVM (SEV-SNP) & AMD SEV-SNP VM with Linux guest OS \\
 \hline 
 \projectname{} & Our \projectname{} system\\
 \hline 
\end{tabular}
\label{tab:variants}
\end{table}

\begin{figure*}[t]
    \centering
    \includegraphics[width=2\columnwidth]{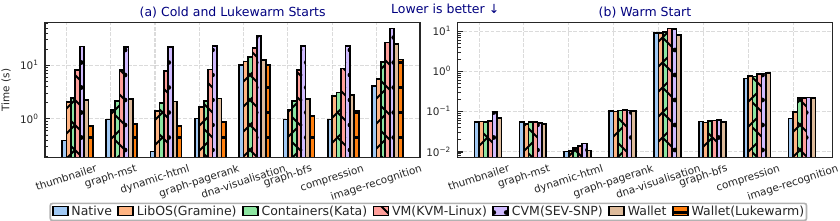}
    % \vspace{-1mm}
    \caption{
    SeBS benchmark end-to-end latency: (a) cold and lukewarm starts, and (b) warm start.
    \textit{
    %(a) \projectname{} achieves faster end-to-end cold start latencies thanks to the optimized function startup.
    %(b) \projectname{} also achieves comparable or even better warm start execution to other baselines on average.
    }}
    \label{fig:sebs:client}
    % \vspace{-4mm}
   % \Description[<short description>]{<long description>}
\end{figure*}

% The \emph{Native} runs the application directly on bare metal.
% The \emph{LibOS (Gramine)} executes the Gramine LibOS on bare metal. 
% The \emph{Container (Kata)} employs the Kata containers runtime with QEMU as its hypervisor. 
% In the \emph{VM (KVM-Linux)} variant, the application operates within a standard VM running Linux as its guest.
% The \emph{CVM (SEV-SNP)} runs a standard AMD SEV-SNP configuration with a Linux guest OS. 
% The \emph{\projectname{}} is our proposed framework.

%Finally, the \emph{\projectname{}} variant represents the application deployed within the \projectname{} framework.
%Unless explicitly noted, we run \projectname{} with memory preallocation optimization enabled.

%\pramod{cite the eurosys paper, and also make it short, just say that evaluation using two workloads, benchmark and then azure traces.}
%\masa{we do not use SeBS-flow, but use  the original SeBS}
\myparagraph{Workloads}
For performance evaluation, we use Python functions from the SeBS benchmark suite~\cite{copik2021sebs}.
\revision{The current \projectname{} prototype lacks external network and \emph{fork()} support; therefore, we exclude ``uploader'', ``crud-api'' (requires external networking), and ``video-processing'' (requires running \emph{ffmpeg} as a standalone process).}
Additionally, we conduct a simulation-based study with Azure Functions production traces~\cite{harvested_serverless, serverless_in_the_wild}.

\subsection{End-to-end Performance}
\label{subsec:performance}
%First, we present end-to-end serverless function performance on the SeBS benchmark suite~\cite{copik2021sebs}.

\myparagraph{Methodology} 
%We run the SeBS server for each variant. % (\autoref{tab:variants}),
%The client is co-located with the function execution environment on the same host.
We run the SeBS server in each variant and co-locate the client on the same host.
%In \projectname{}, both the SeBS server and the client run in the guest OS, and functions are executed as trustlets.
%We measure (1) end-to-end latency and (2) invocation latency, the duration from the start of the client request to the beginning of the function execution.
% We extend the SeBS benchmark suite to run serverless functions with \projectname{} using \projectname{}'s python library.
\projectname{} does not include the time of measurement calculation for a fair comparison with the other baselines.
% We present the measurement time analysis in \autoref{sec:eval:analysis}.
We run each experiment \emph{five} times and report the average.

\myparagraph{End-to-end latency}
\autoref{fig:sebs:client} presents the end-to-end client latency of each SeBS function for each variant.
In the cold start case, \projectname{} (Lukewarm) achieves 93.44\% speedup over CVM (SEV-SNP) on average, and is 83.99\%, 51.04\%, 32.85\% faster than VM (KVM-Linux), Containers (Kata), and LibOS (Gramine).
Even in the ``cold''est case (i.e., no loaded zygote), \projectname{} is on average 85.10\% faster than the CVM (SEV-SNP) baseline and 63.62\% faster than the VM.
When compared to the Containers (Kata), and LibOS (Gramine), \projectname{} incurs 11.25\% and 52.57\% performance overhead, respectively.
%These performance gains mainly come from the \projectname{}'s efficient function startup (\autoref{sec:eval:analysis:invocation}).
%
%client_time (warm start) - Wallet Comparison:
%  Wallet is 20.81% slower than Native                 
%  Wallet is 15.63% slower than LibOS(Gramine)
%  Wallet is 1.04% faster than Containers(Kata)                  
%  Wallet is 7.53% faster than VM(KVM-Linux)                
%  Wallet is 14.97% faster than CVM(SEV-SNP)
In the warm start case, the performance difference among variants becomes smaller as the existing function environment is reused. 
Still, \projectname{} is 14.97\%, 7.53\% and 1.04\% faster than the CVM, VM and Containers (Kata) case respectively while it shows a 15.63\% performance overhead relative the LibOS (Gramine).
%The run-to-completion and in-memory file system based execution of \projectname{} minimizes context switch overhead, thus achieving the high execution performance (\autoref{sec:eval:analysis:execution}).
\projectname{}'s performance gains mainly stem from its efficient function startup, run-to-completion execution model that minimizes VMEXITs, and in-memory filesystem.
\projectname{} exhibits higher latency if the input/output size is large due to data transfer overhead between different VMPL levels (e.g., image-recognition has 100MB input) (\autoref{sec:eval:analysis}).
\myparagraph{Invocation latency}
% The cold-start overhead primarily affects the invocation latency.
\autoref{fig:sebs:invocation_latency_cdf} shows the cumulative distribution function (CDF) of invocation latency under cold starts, which primarily affect the invocation latency.
\projectname{} (Lukewarm) has p50/p99 (50th and 99th percentile) latencies of 0.019/1.06 s.
It achieves ms-scale p50 startup latency, comparable to the latest (non-confidential) fork-based serverless systems~\cite{seuss,catalyzer,sock,du_xpusim_2022,li_pagurus_2022}.
On the other hand, \projectname{} (cold) has 1.57/12.34 s latency, which is at the levels of Containers (1.93/2.10 s) and much lower than the CVM (21.25/22.10 s) and the VM (7.5/8.0 s) variants.
It also exhibits a larger standard deviation (3.51 s), %compared to the Containers (0.04 s) or Gramine LibOS (0.06 s)
which 
originates from the time needed to load the zygote and the variably-sized functions (\autoref{sec:eval:analysis}).
% XXX: why other baselins does not have this issues? => using fs?

\subsection{Performance Analysis}
\label{sec:eval:analysis}
%\textbf{RQ2:} \textit{What factors affect the performance of \projectname{}?}
% To understand the contribution of \projectname{}'s design choices to performance, this section presents an in-depth analysis of function invocation and execution.
%We split end-to-end execution into two phases, (1) invocation phase and (2) execution phase, and present the breakdown analysis. 
%In addition, we also present the analysis of memory preallocation and CoW mechanism of \projectname{}.

%\subsubsection{Breadkdo}
%\label{sec:eval:analysis:invocation}

%The function invocation includes two stages: (1) booting the environment (for the cold start), and (2) initializing the runtime.

% result: ./Benchmarks/Boottimes/plot.py results.txt
% Walle (init):  Total: 12358.98 ms
% CVM (SEV-SNP): Total:  8307.31 ms (12358.98-8307.31=4051.67)
% Wallet (cold): Total:  2388.14 ms
% Wallet (warm): Total:    10.27 ms % XXX: now lukewarm
% Wallet (hot):  Total:     9.62 ms % XXX: now warm
\myparagraph{Boot time analysis}
\autoref{fig:mot:boot} presents the boot time breakdown. 
For VM-based variants, %(Containers (Kata), VM, CVM, and \projectname{}),[
the majority of the boot time is consumed by VMM (QEMU) initialization and  guest OS startup.
The CVM incurs additional overhead due to extra management tasks (e.g., measurement of initial state, memory validation)~\cite{misono2024confidential}.
Although \projectname{}'s environment setup takes 4.05 s longer than the CVM (without memory preallocation (\autoref{sec:eval:analysis:optimization})) because of the added setup time for the monitor, \projectname{}'s cold start only takes 2.38 s by eliminating the CVM initialization time.
\projectname{}'s lukewarm start further optimizes the boot time by employing a fork-based start (10.3 ms).

% ./Benchmarks/SeBS_analysis/plot.py
\begin{figure*}[t]
\centering
\begin{subfigure}{0.33\textwidth}
    \centering
    \includegraphics[width=\textwidth]{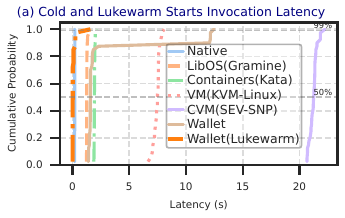}
    \vspace{-0.6cm}
    %\caption{Function invocation CDF.}
    \phantomcaption
    \label{fig:sebs:invocation_latency_cdf}
%    \Description[<short description>]{<long description>}
\end{subfigure}
\hfill
% ./Benchmarks/SeBS_analysis/plot.py
\begin{subfigure}{0.33\textwidth}
    \centering
    \includegraphics[width=\textwidth]{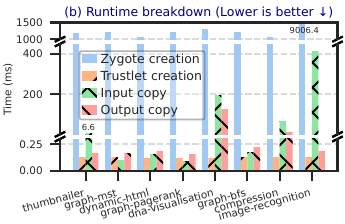}
    \vspace{-0.6cm}
    %\caption{Runtime processing breakdown.}
    \phantomcaption
    \label{fig:sebs:init:breakdown}
%   \Description[<short description>]{<long description>}
\end{subfigure}
\hfill
% ./Benchmarks/SeBS_analysis/plot_memory.py
\begin{subfigure}{0.33\textwidth}
    \centering
    \includegraphics[width=\textwidth]{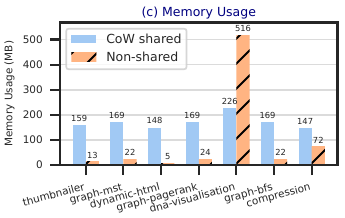}
    \vspace{-0.6cm}
    %\caption{Memory utilization.}
    \phantomcaption
    \label{fig:sebs:mem:cow}
 %  \Description[<short description>]{<long description>}
\end{subfigure}
% \vspace{-0.4cm}
\caption{Performance analysis: (a) invocation latencies, (b) \projectname{}'s runtime breakdown, and (c) \projectname{}'s memory usage.}
% \vspace{-6mm}
\end{figure*}

\myparagraph{Runtime processing cost}
Further, we analyze the runtime processing of \projectname{}, focusing on zygote and trustlet creation, and input/output transfer. % during the function execution.
%\autoref{fig:sebs:init:breakdown} presents the breakdown of the zygote creation, trustlet creation, and input and output copy time during the function execution.
% \autoref{fig:sebs:init:breakdown} shows the results.
Each process involves transferring data between the guest OS and the trusted monitor, followed by updating the VMPL level and the trustlet's page tables. % to configure access rights.
As \autoref{fig:sebs:init:breakdown} illustrates, the zygote creation is responsible for most of the runtime initialization. %, as it loads the base LibOS image from the guest OS into the trusted monitor memory.
The current prototype bundles required libraries in the base image (e.g., numpy, pytorch), increasing the zygote size, ranging from 60 MB to 691 MB (image-recognition) in SeBS.
However, zygote creation occurs only during cold starts, and the CoW mechanism enables efficient memory sharing among trustlets using the same zygote image (\autoref{sec:eval:analysis:memory}).

On the other hand, the trustlet creation and data transfers' duration depends on the function and I/O size.
Each SeBS function is less than 4 KB. Thus, the trustlet creation requires $<0.2$ ms.
However, larger data transfers, such as in dna-visualisation 
(112 MB output), take $\sim122$ ms.
% (4.1 MB input, 112 MB output), last 192 ms and 122 ms, respectively.
If the data fits within a 4K page, the copying time is $<0.1$ ms (e.g., graph-mst and graph-pagerank).

\myparagraph{Measurement cost}
\autoref{fig:mot:attestation} shows the breakdown of the time taken to calculate the SHA512 measurement of an empty Python function.
The calculation time is proportional to the data size.
%In the cold start case, the measurements include those of the trusted monitor, as well as the zygote (58 MB), which takes around 2.7 s.
In the cold start case, the measurements include those of the trusted monitor, as well as the zygote (60 MB), which takes around 1.1 s.
The current prototype does not support CPU acceleration for SHA, justifying the lower performance.
However, for the lukewarm start, \projectname{} only needs to measure the function, input, and output (typically less than 4KB), dropping the calculation time below $0.5$ ms.

\subsubsection{Effectiveness of Optimizations}
\label{sec:eval:analysis:optimization}
%We now present an ablation study of \projectname{}'s two key optimizations: memory preallocation and copy-on-write (CoW).
%Specifically, we execute SeBS benchmarks with and without these optimizations and compare the results.
%We now present an ablation study of \projectname{}'s optimizations by comparing SeBS results with and without these optimizations.

% XXX: this is to adjust the section name position when having myparagraph directly after the section name
\myparagraph{Memory preallocation} 
%\textbf{RQ2:} \textit{What is the impact of \projectname{}’s memory preallocation optimization on boot time and function invocation?}
%We analyze the effect of memory preallocation on the end-to-end execution.
Memory validation% (via the pvalidate instruction) in the SEV-SNP VM
~\cite{sev-snp-abi} is expensive as it involves VMEXIT for state updates.
In our setup, validating a 4K page takes 24 \textmu s (6ms/MB).
Although this cost is an one-time overhead, it affects the initial creation time of zygotes and trustlets.
Memory preallocation removes this validation cost at runtime in exchange for an increased bootup time.
%\autoref{fig:opt:prealloc:exec} shows the end-to-end execution latency  with and without memory preallocation.
For the SeBS benchmark, the preallocation improves the performance by 49.22\% in the cold start and 4.72\% in the warm start case on average, over the non-preallocation version.
% XXX: the reason for speedup of warm-start case: presumably even warm start up case some new allocation happens. I assume that if we keep a function warm, then eventually  all pages get validated and there would be no performance difference between prealloc and w/o prealloc
%On the other hand, \autoref{fig:opt:prealloc:boot} shows the impact on the boot time when preallocates 16 GB memory.
%Although this increases the boot time by 238 s, it removes the page-validation overhead during the runtime
However, preallocating 16 GB of memory increases the boot time by 238 s, highlighting the trade-off.
% However, this removes the page-validation overhead during the runtime.
% XXX: current implementation of preallocation seems slow as pvalidating 16GB would take ~100s based on the microbenchmark result. I guess this is because when validating (iiuc) the monitor creating a temporal page table entry, which is costly. 

%\masa{low-level analysis: (1) microbenchmark of rmpadjust/pvalidate, showing the cost of memory allocation}
%\masa{(2) report how many rmpajudst/pvalidate happens at runtime}

% result: Benchmarks/SeBS_analysis/plot_cow.py
%Performance Comparison Table:
%Benchmark                 Cold (w/o) Cold (w/)  Cold Impr. Hot (w/o)  Hot (w/)   Hot Impr. 
%--------------------------------------------------------------------------------
%thumbnailer               2.9205s    2.8933s    0.93%      0.0729s    0.0722s    0.97%     
%graph-mst                 2.6082s    2.5038s    4.01%      0.0520s    0.0505s    2.91%     
%dynamic-html              2.2875s    2.2495s    1.66%      0.0103s    0.0103s    0.77%     
%graph-pagerank            2.6228s    2.5695s    2.03%      0.1026s    0.1041s    -1.44%    
%dna-visualisation         16.4832s   13.7907s   16.33%     11.9098s   9.4099s    20.99%    
%graph-bfs                 2.6234s    2.4814s    5.41%      0.0572s    0.0545s    4.66%     
%compression               3.6142s    3.4820s    3.66%      0.9831s    0.9551s    2.85%     
%--------------------------------------------------------------------------------
%Geometric Mean                                         3.28%                           N/A  

\myparagraph{Copy-on-Write (CoW)}
%To instantiate a trustlet from a zygote, \projectname{} employs CoW, which removes the full memory copy of zygote memory and thus reduces the trustlet creation time.
We further investigate the impact of CoW.
%\autoref{fig:sebs:cow_perf} shows the affect of CoW to the end-to-end latency of SeBS benchmarks.
Performance-wise, CoW is crucial for achieving rapid trustlet creation and, thus, fast lukewarm starts (\autoref{fig:sebs:init:breakdown}).
CoW reduces the average trustlet creation time of SeBS functions from 66 ms to 0.11 ms. % Trustelet creation time: from the breadkwon analysis
However, it incurs additional overhead during execution due to page faults. 
This overhead becomes noticeable mostly in the warm starts, but still remains almost negligible ($\sim$1\%), promoting CoW's adoption. %, conclude that CoW is overall beneficial for \projectname{}.  %indicating that the speedup in initialization time outperforms runtime overhead caused by page faults.

\begin{figure}
    \centering
    \includegraphics[width=0.45\textwidth]{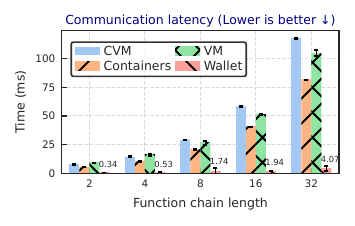}
    % \vspace{-4mm}
    \caption{Latencies of function chains of different lengths.}
    % \vspace{-3mm}
    \label{fig:eval:comm_analysis}
\end{figure}

\subsection{Resource Efficiency}
\label{sec:eval:analysis:memory}
% Next, we analyze \projectname{} regarding memory efficiency.

% Wallet memory analysts: ./Benchmarks/SeBS_analysis/plot_memory.py
%Overall Memory Usage:
%Average CoW Memory Usage: 169.91 MB
%Average No-CoW Memory Usage: 96.29 MB
%On average, CoW uses 73.62 MB more memory (76.46% more)
\myparagraph{Memory footprint}
We evaluate the memory footprint of \projectname{} while running SeBS functions.
%We extend the monitor's memory allocator to record the statistics of memory allocation and report the peak memory usage during the execution of a function.
\autoref{fig:sebs:mem:cow} presents the result.
``CoW shared'' represents the memory shared with CoW among trustlets and their base zygote, whereas ``Non-shared'' denotes the memory exclusively allocated for a trustlet during execution.
Although the amount of allocated memory (``Non-shared'') varies depending on the workload, on average, trustlets share 170 MB of memory with the base zygote, reducing the total system memory consumption.

\myparagraph{Function density}
Memory sharing enables \projectname{} to pack more functions per node.
We analyze the memory consumption for increasing numbers of concurrently-running empty Python functions, shown in \autoref{fig:mot:function_density}.
% As shown in \autoref{fig:mot:function_density}, we analyze the memory consumption required to host empty Python functions with increasing the number of functions.
We enable kernel same-page merging (KSM)~\cite{ksm} on the host to de-duplicate identical memory pages.
%KSM effectively reduces memory usage for Kata and VM, but not for CVM and \projectname{}, as their memory is encrypted with unique keys.
% While the KSM is effective for Kata and VM, it is not effective for CVM and \projectname{} since their memory regions are encrypted with unique keys.
% For \projectname{}, we assess the required memory consumption by multiplying the amount of non-CoW-shared memory used by a single function (60KB) by the total number of functions, in addition to the base shared memory (147MB).
For \projectname{}, the total memory is calculated as the base zygote shared memory (147 MB) plus 60 KB of non-CoW-shared memory per function.
As KSM cannot deduplicate encrypted memory, CVM memory consumption is proportional to the number of functions, leading to a higher per-function memory overhead.
%Our results show that the CVM memory consumption scales proportionally to the number of functions, having a much higher per-function memory overhead.
Further, the number of CVMs in a single node is limited to the number of encryption keys (509 in our environment).
In contrast, \projectname{} achieves significantly lower memory usage, realizing high function density.

\subsection{Communication Analysis}
\label{subsec:networking-eval}
%\pramod{JUST one plot for function chaining with changing the number of functions}
\myparagraph{Methodology} 
To evaluate \projectname{}'s inter-function communication, %we measure the overhead of invoking multiple functions in a chain. Specifically, we vary the number of functions in the chain and measure the total communication time.
we measure the communication time of invoking multiple functions in a chain by varying the data size and the number of functions in the chain.
All functions are deployed on the same host.
In \projectname{}, each function creates its communication endpoint for input and output, while other baselines use TCP networking.

\myparagraph{Communication latency}
First, we show the communication latency between two functions under various data sizes.
\autoref{fig:mot:ipc} shows the result.
This result indicates that \projectname{} achieves lower latency than CVMs (1.4-27$\times$), as well as traditional VMs and Containers (Kata), across message sizes, thanks to the zero-copy-based data objects.

\myparagraph{Latencies of function chains}
% \autoref{fig:eval:comm_analysis}
Next, we show the latencies of function chains.
As shown in \autoref{fig:eval:comm_analysis}, \projectname{} outperforms all baselines across all chain lengths, achieving 16.7-30.2$\times$, 11.8-20.9$\times$, and 15.2-30.1$\times$ lower communication latency than CVMs, Containers, and traditional VMs, respectively. \projectname{}'s performance advantage grows with chain length, demonstrating the efficiency of its data-centric I/O for inter-function communication compared to TCP-based networking used by other approaches, highlighting \projectname{}'s suitability for complex workloads with function chaining.

%\masa{possible addition: latencies of a two-function chain invocation under various data sizes (pheromone fig 11)}

\begin{figure*}[t]
\centering
\begin{subfigure}{0.33\textwidth}
    \centering
    \includegraphics[width=\textwidth]{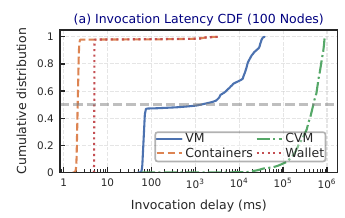}
    \vspace{-0.7cm}
    % \caption{Invocation latency CDF.}
    \phantomcaption
    \label{fig:eval:scheduling:delay}
%    \Description[<short description>]{<long description>}
\end{subfigure}
\hfill
\begin{subfigure}{0.33\textwidth}
    \centering
    \includegraphics[width=\textwidth]{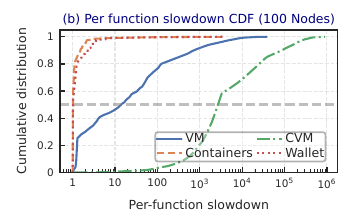}
    \vspace{-0.7cm}
    % \caption{Per-function slowdown CDF.}
    \phantomcaption
    \label{fig:eval:scheduling:slowdown}
 %   \Description[<short description>]{<long description>}
\end{subfigure}
\hfill
\begin{subfigure}{0.33\textwidth}
    \centering
    \includegraphics[width=\textwidth]{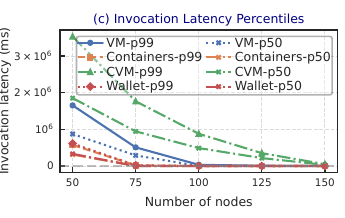}
    \vspace{-0.7cm}
    % \caption{Invocation latency with varying nodes.}
    \phantomcaption
    \label{fig:eval:scheduling:node}
 %   \Description[<short description>]{<long description>}
\end{subfigure}
% \vspace{-0.4cm}
\caption{Scale-out results using Azure Functions production traces~\cite{harvested_serverless, serverless_in_the_wild}.}
% \vspace{-6mm}
\end{figure*}

\subsection{Scale-out Performance}
\label{subsec:traces}
% Lastly, we evaluate \projectname{}'s scale-out performance with a simulation-based study to demonstrate its effectiveness under a production-like, large-scale environment.

\myparagraph{Methodology}
% We develop a simulator to model real-world serverless deployments. It consists of a central scheduler and multiple nodes, each with fixed execution slots and LRU-based caches to keep warm functions.
We develop a simulator modeling real-world serverless deployments with a central scheduler and multiple nodes, each having fixed execution slots and LRU caches for warm functions.
% We develop a simulator to resemble real-world severless function deployments.
% It consists of a central scheduler and a set of nodes.
% Each node has a fixed number of execution slots and LRU-based caches to keep warm functions. %\teofil{Maybe it was done due to space constraints but mentioning LRU in the same sentence as the the cache sloots feels a bit weird}
% The scheduler processes trace data by arrival time, prioritizing nodes that already cache the requested function.
The scheduler processes trace data chronologically, prioritizing nodes already caching the requested function.
If none are available, it assigns requests to free nodes or queues them.
% it assigns the request to a free node or delays it until one becomes available.
% If there is no such node, it sends the request to an available
%\teofil{technically assigns it to a node that has a free exec slot, not necesseraly an idle one. I guess it depends how we want to present the nodes here, i.e. if we present them as 1 func/node}
% node, if any; otherwise, the scheduler delays the request until one node becomes free. 
%\teofil{Maybe explain how we treat lukewarm: The lukewarm boot type happens only in the Wallet case. In our simulation, a lukewarm boot happens when a function runs on a node that has in its cache another function from the same aplication as the current function. Thus we simulate the zygote reutilization under the assumption that functions from the same application share a zygote.}
Function execution duration gets increased based on its boot type (cold/lukewarm/warm), using timings sampled from our microbenchmarks (\autoref{fig:mot:boot}).
% When a function is scheduled on a node, its \textit{duration} is increased depending on its boot type (cold/lukewarm/warm), which are sampled from the distributions obtained by our microbenchmarks (\autoref{fig:mot:boot}).
The lukewarm boot, unique to \projectname{}, occurs when a node caches another function from the same application, simulating zygote sharing.
We use Azure Functions production traces~\cite{harvested_serverless, serverless_in_the_wild} sampled with InVitro~\cite{ustiugov:invitro}, comprising 4k functions with $\sim$4.1 million invocations over 30 minutes.

\myparagraph{Invocation latency and per-function slowdown}
We run the simulator with 100 nodes, each having 32 execution slots and a cache size of 32.
% \autoref{fig:eval:scheduling:delay} shows the CDF of invocation latency.
\autoref{fig:eval:scheduling:delay} shows the CDF of invocation latency: at p50/p99, CVM and VM experience delays of 489/881 s and 1.3/33 s, respectively, while \projectname{} achieves just 5 ms/1.5 s.
% At the p50/p99 points, the CVM and the VM have 489/881 s and 1.3/33 s scheduling delay, respectively.
% In contrast, the \projectname{} maintains 5 ms/1.5 s latencies.
\autoref{fig:eval:scheduling:slowdown} reports the per-function slowdown, which increases as function execution time decreases and invocation latency dominates. 
At p50/p99, CVM slowdowns reach 2,682/391,257, VM slowdowns are 13.98/8,658, while, in contrast, \projectname{}'s slowdowns are merely 1.02/7.06.
\myparagraph{Node scalability}
We also evaluate the invocation latency for various numbers of nodes.
\autoref{fig:eval:scheduling:node} shows that, compared to the other baselines, \projectname{} achieves lower latency with a smaller number of nodes.
For example, \projectname{} has 5 ms p99 latency with 100 nodes, while, even with 150 nodes, the CVM's p99 latency is 50 s, and the VM's is 12 s.
These results highlight \projectname{}'s effectiveness in a large-scale environment.

\section{Related Work}

\myparagraph{Serverless computing}
Serverless computing has transformed cloud deployment. Open-source frameworks (OpenWhisk~\cite{openwhisk}, OpenLambda~\cite{openlambda}, OpenFaaS~\cite{openfaas}) and major cloud providers (AWS Lambda~\cite{aws-lambda}, Azure Functions~\cite{azure-functions}, Google Cloud Functions~\cite{google-cloud-functions}) offer widely used architectures. Recent research~\cite{boki, xfaas, nightcore, catalyzer, harvested_serverless, icebreaker, sigmaos, sand, sock, faasm, seuss} mainly focuses on improving serverless isolation, performance and efficiency,
%(e.g., Nightcore~\cite{nightcore} for container isolation and low latency, Boki~\cite{boki} for stateful applications). 
while \projectname{} targets the security challenges and leverages TEEs to form a lightweight confidential computing system for secure serverless deployments.

\myparagraph{Confidential computing}
Confidential computing, based on hardware TEEs~\cite{sev, sev-snp, tdx, cca, trustzone, sgx, lee_keystone_2020}, is widely adopted for protecting data and code in untrusted clouds. 
% TEEs include process-based (e.g., Intel SGX~\cite{sgx}) and VM-based (e.g., AMD SEV-SNP~\cite{sev-snp}, Intel TDX~\cite{tdx}, ARM CCA~\cite{cca}) types.
TEE technologies can be broadly categorized into process-based (e.g., Intel SGX~\cite{sgx}) and VM-based TEEs (e.g., AMD SEV-SNP~\cite{sev-snp}, Intel TDX~\cite{tdx}, ARM CCA~\cite{cca}). 
Extensive recent research~\cite{park_nestedenclave_2020, gu_lightenclave_2022, zhao_resusableenc_2023, veil_ahmad_2024, sev_svsm, kuvaiskii2024gramine, ge_hecate_2022, dataenclave, domainisolation, severifast, acctee, twine, sponsshields, glamdring, cheri-tree, capstone, li_bifrost_2023, wang_nestedsgx_2025, zhang_erebor_2025, anchor2024, tnic2025, bailleu_speicher_2019} aims to minimize TCB, and provide intra-VM isolation, secure data at rest, among others.

Specifically, Veil~\cite{veil_ahmad_2024},  and NestedSGX~\cite{wang_nestedsgx_2025} deploy secure services in CVMs using SEV-SNP VMPL.
%SVSM~\cite{sev_svsm, coconut_svsm} is a privileged software providing secure services to the guest OS in CVM with SEV-SNP VMPL or TD-Partitionig.
Erebor~\cite{zhang_erebor_2025} adopts  intra-kernel privilege isolation to provide sandboxing in CVMs.
%Hecate~\cite{ge_hecate_2022} implements an L1 hypervisor shielding nested VMs from the untrusted L0 hypervisor. 
Unlike these, \projectname{} targets serverless deployments, and leverages CVM partitioning for minimal TCB, fast boot, and efficient function communication via shared memory.

\myparagraph{Confidential serverless computing}
Many research works aim for secure serverless computing architectures using confidential computing~\cite{se-lambda, s-faas, pluginenclaves, serverlessCoCo, cryonics, trust-more-serverless, penglai, clemmys, sandbox-enclaves-serverless}. 
ServerlessCoCo~\cite{serverlessCoCo} comprehensively analyzes overheads in CVM-based serverless deployments.
Plugin Enclaves~\cite{pluginenclaves} proposes a hardware-based approach with Intel SGX for efficient confidential serverless computing. Cryonics~\cite{cryonics} reduces startup times using snapshot-based SGX enclaves.  
%using attested and encrypted container images. 
In contrast, \projectname{} materializes confidential serverless computing by providing a CVM-based lightweight system with easily attestable and deployable trusted processes.

\revision{
Concurrently with our work, COFUNC~\cite{shi_cofunc_2025} proposes a library OS-based confidential containers tailored for serverless functions.
COFUNC employs a microkernel architecture, running multiple containers with the same CVM while enforcing isolation.
Contrary to \sysname, COFUNC offloads container management tasks to the serverless runtime that runs on the host. On the other hand, \sysname allows for the reuse of the existing OS and serverless framework within the CVM, leveraging its nested confidential execution architecture.
}

\myparagraph{Lightweight virtualization}
Lightweight virtualization techniques~\cite{firecracker, kata, unikraft, gramine1, gramine2, kuvaiskii2024gramine, OSv, LightVM, unikernels} aim to reduce the VM overheads while maintaining isolation.
%Firecracker~\cite{firecracker} introduces microVMs for serverless workloads with minimal boot times. Kata Containers~\cite{kata} deploy lightweight VMs with container-level isolation in standard VMs. 
Firecracker~\cite{firecracker} introduces microVMs for serverless workloads. 
Kata Containers~\cite{kata} deploy lightweight VMs for container applications, whereas 
Unikraft~\cite{unikraft} creates specialized unikernels for applications.
Gramine LibOS~\cite{gramine1, gramine2, kuvaiskii2024gramine} runs unmodified apps in SGX enclaves and TDX VMs\revision{~\cite{kuvaiskii2024gramine}}.
%\revision{In both cases, Gramine is a single-tenant runtime occupying the entire TEE, requiring multiple instances to run multiple functions. } 
On the other hand, \projectname{} leverages these concepts with a minimal \revision{Gramine} LibOS \revision{instance} for \revision{each of} its trusted processes, enabling low boot and attestation times, and seamless application deployment.
\revision{Unlike Gramine-TDX, \projectname{}'s LibOS is only running in a certain privilege level within the CVM and does not have system-wide privileges. Further, in contrast to Gramine's single-tenant model with its existing backends, \projectname{} provides a framework for managing multiple instances within the same CVM, enabling multitenancy and optimized inter-function communication via data objects among trusted processes while maintaining function instance protection.}

\myparagraph{Networking/IO for confidential computing}
%\pramod{add papers like shieldbox, rkt-io, avocado, bitfrost, etc.}
TEE networking is typically slow due to the added overhead of encryption and data copying with multiple context switches~\cite{li_bifrost_2023,misono2024confidential,weisse_hotcalls_2017}
Several works optimize their processing overhead with polling at the cost of increased CPU usage~\cite{jorg_rktio_2021,trach_shieldbox_20018,bailleu_avocado_2021, bailleu_speicher_2019}.
%Bifrost~\cite{li_bifrost_2023} optimizes CVM network processing by removing redundant bounce buffer copying while preserving protocol safety guarantees.
Compared to them, \projectname{} achieves efficient inter-function communication through its data-centric shared-memory-based I/O architecture while ensuring isolation.

\section{Conclusion}
\label{sec:conclusion}
% \dimitris{WIP}
In this paper, we present \projectname{}, a lightweight confidential serverless computing system. % framework tailored for serverless workloads in untrusted cloud environments. 
\projectname{} leverages intra-CVM partitioning mechanisms and efficiently consolidates security-critical functionalities into a compact, privileged trusted monitor, resulting in a small TCB. 
Through its zygote mechanism, \projectname{} achieves fast boot times while optimizing memory management and reducing the communication cost between serverless functions via its data-centric I/O architecture. Further, it provides a formally verified differential attestation mechanism to ensure end-to-end trust with reduced attestation times. 
Overall, \projectname{} introduces low-performance overheads while exposing a minimal attack surface, making it ideal for security- and latency-sensitive serverless deployments.

\myparagraph{Artifact} \projectname{} is publicly available at \url{https://github.com/TUM-DSE/Wallet-VMPL} along with its entire experimental setup.

% \newpage
\subsection*{Acknowledgments}
We thank our shepherd, Prof. Anurag Khandelwal, and the anonymous reviewers for their helpful comments.
We also thank Prof. Nuno Santos for valuable discussion and feedback on this work.

This work was supported in part by an ERC Starting Grant (ID: 101077577) and the Chips Joint Undertaking (JU), European Union (EU) HORIZON-JU-IA, under grant agreement No. 101140087 (SMARTY), the Intel Trustworthy Data Center of the Future (TDCoF), and Google Research Grants. 
The authors acknowledge the financial support by the Federal Ministry of Research, Technology and Space of Germany in the programme of “Souverän. Digital. Vernetzt.”. Joint project 6G-life, project identification number: 16KISK002.

\bibliographystyle{plainurl}
\bibliography{main}

% \clearpage
% \appendix
% \input{sections/appendix.tex}
%-------------------------------------------------------------------------------

% previous text
% \newpage
% \input{pre.tex}

%%%%%%%%%%%%%%%%%%%%%%%%%%%%%%%%%%%%%%%%%%%%%%%%%%%%%%%%%%%%%%%%%%%%%%%%%%%%%%%%
\end{document}
%%%%%%%%%%%%%%%%%%%%%%%%%%%%%%%%%%%%%%%%%%%%%%%%%%%%%%%%%%%%%%%%%%%%%%%%%%%%%%%%

%%  LocalWords:  endnotes includegraphics fread ptr nobj noindent
%%  LocalWords:  pdflatex acks